\documentclass[fleqn,12pt]{article}
\usepackage{amsfonts,epsfig}
\usepackage{latexsym,amssymb}

\newcommand{\ft}[2]{{\textstyle\frac{#1}{#2}}}
\def\tilde{\widetilde}

\def\1bar{1\hskip -.275cm -}
\def\2bar{2\hskip -.275cm -}
\def\3bar{3\hskip -.275cm -}

\newsavebox{\uuunit}

\makeatletter \@addtoreset{equation}{section} \makeatother

\def\bfone{\relax{\rm 1\kern-.35em 1}}

\def\bfone{\relax{\rm 1\kern-.35em 1}}
 
\newcommand{\nc}{\newcommand}

\newcommand{\Ge}{\epsilon}

\newcommand{\bbR}{{\mathbb{R}}}
\newcommand{\bbZ}{{\mathbb{Z}}}

\newcommand{\CO}{{\cal O}}

\newcommand{\la}{\label}

\newcommand{\Ref}[1]{(\ref{#1})}

\newcommand{\dd}{\partial}

\newcommand{\equ}{\!=\!}
\newcommand{\pls}{\!+\!}
\newcommand{\plss}{\!\!+\!\!}
\newcommand{\mis}{\!-\!}
\newcommand{\miss}{\!\!-\!\!}
\providecommand{\mathring}[1]{{\stackrel{\mbox{\tiny o}}{#1} }}
\newcommand{\mathon}{\mathversion{bold}}
\newcommand{\mathoff}{\mathversion{normal}}

\nc{\be}{\begin{equation}} \nc{\ee}{\end{equation}}
\nc{\bea}{\begin{eqnarray}} \nc{\eea}{\end{eqnarray}}
\newcommand{\ben}{\begin{displaymath}}
\newcommand{\een}{\end{displaymath}}
\nc{\dalpha}{\dot{\alpha}} \nc{\dbeta}{\dot{\beta}}
\nc{\nn}{\nonumber} \nc{\non}{\nonumber\\}
\begin{document}

\thispagestyle{empty}

\begin{flushright}
ITF-2002/28\\
SPIN-2002/18\\
{\tt hep-th/0206247}
\end{flushright}
\renewcommand{\thefootnote}{\fnsymbol{footnote}}

\vspace*{0.1ex}

\bigskip\bigskip

\begin{center}
{\bf\Large Supergravity Duals of Matrix String Theory}
\bigskip\bigskip\medskip

{\bf Jose F.~Morales and Henning Samtleben \medskip\\ }
{\em Institute for Theoretical Physics} \,and\, {\em Spinoza Institute},\\ 
{\em Utrecht University, Postbus 80.195, 3508 TD Utrecht, The Netherlands}
\smallskip

{\small morales@phys.uu.nl, h.samtleben@phys.uu.nl}

\end{center}

\setcounter{footnote}{0}
\bigskip
\medskip

\begin{abstract}

We study holographic duals of type II and heterotic matrix string
theories described by warped $AdS_3$ supergravities.  By explicitly
solving the linearized equations of motion around near horizon
D-string geometries, we determine the spectrum of Kaluza-Klein
primaries for type I, II supergravities on warped $AdS_3\times
S^7$. The results match those coming from the dual two-dimensional
gauge theories living on the D-string worldvolumes. We briefly discuss
the connections with the ${\cal N}=(8,8)$, ${\cal N}=(8,0)$ orbifold
superconformal field theories to which type IIB/heterotic matrix
strings flow in the infrared.  In particular, we associate the
dimension $(h,\bar{h})=(\ft32,\ft32)$ twisted operator which brings
the matrix string theories out from the conformal point
$(\bbR^8)^N/S_N$ with the dilaton profile in the supergravity
background.

The familiar dictionary between masses and ``scaling'' dimensions of
field and operators are modified by the presence of non-trivial warp
factors and running dilatons. These modifications are worked out for
the general case of domain wall/QFT correspondences between
supergravities on warped $AdS_{d+1}\times S^q$ geometries and super
Yang-Mills theories with $16$ supercharges.

\end{abstract}

\renewcommand{\thefootnote}{\arabic{footnote}}
\vfill

\bigskip

\leftline{{June 2002}}

\setcounter{footnote}{0}
\newpage

\tableofcontents

\section{Introduction}

Soon after Maldacena's proposal \cite{maldacena} for a holographic
correspondence between string theory on anti-de Sitter spaces and
conformal field theories living at the $AdS$ boundaries, this
correspondence was extended to more general string backgrounds and
non-conformal field theories.  In~\cite{Itzhaki:1998dd}, a proposal
relating string theory on near horizon Dp-brane geometries to
$d=p\pls1$ dimensional super Yang-Mills (SYM) theories with sixteen
supercharges was put forward. These so called domain wall/QFT
dualities, later developed in
\cite{Boonstra:1998mp,Behrndt:1999mk,Gherghetta:2001iv}, provide
the simplest setting in which general ideas of holography can be
tested in more interesting situations involving non-trivial warped
geometries and running dilatons.

Despite its obvious interest, very few it is known at present
about these correspondences. Various aspects were discussed in
\cite{Boonstra:1998mp}; in particular, a series of potential dual
gauged supergravities in various dimensions has been proposed which
have not yet been systematically explored.
In~\cite{Gherghetta:2001iv}, two-point correlations functions for
currents and stress energy tensors have been discussed with special
emphasis on the cases $d=5, 6$ and related brane-world scenarios. The
aim of this paper is a systematic analysis of the spectrum of
Kaluza-Klein (KK) supergravity harmonics and primary operators in the
dual gauge theories for $d=2$. This case is a particularly rich
setting for a domain wall/QFT correspondence. It relates string theory
on certain warped $AdS_3\times S^7$ backgrounds to two-dimensional
fundamental string or gauge theories, depending on whether we study
systems of fundamental or D-strings. Fundamental string solutions are
common to all five ten-dimensional string theories. The study of the
domain wall/QFT correspondence in this case is particularly
interesting since the string background is free of RR fields. The case
of D-strings is particular to type IIB and type I theory and will be
the main subject of our investigation. The dual gauge theories in these
cases correspond to the matrix string models
\cite{Dijkgraaf:1997vv,Banks:1997it} proposed as non-perturbative
definitions of type IIA and heterotic string theories. The
correspondences under study here in principle provide a supergravity
description of this physics (see \cite{Johnson} for early discussions
on these ideas).

In this paper, we study supergravity duals of both matrix string
descriptions. The relevant geometries involve warped $AdS_3\times S^7$
spaces and running dilatons. By means of an harmonic analysis of the
linearized equation of motions around the string backgrounds we
determine the spectrum of primary fields in the corresponding $AdS_3$
supergravities. The results are shown to be in agreement with the dual
description in terms of primary operators in the gauge theory. An
important difference with pure $AdS$/CFT correspondences is the fact
that the conformal group is no longer part of the background isometry
group of this background which is rather given by the semidirect
product of $ISO(1,1)\times SO(8)$ (two-dimensional Poincar\'e group
and sphere isometries) with $16$ supercharges. However, the full
two-dimensional conformal algebra is realized as asymptotic conformal
Killing isometries of the string background.  In addition, these gauge
theories are known to have an interesting IR dynamics governed by
exactly solvable superconformal field theories (SCFTs) which can be
thought of as second quantized type II or heterotic strings moving on
$(\bbR^8)^N/S_N$~\cite{Dijkgraaf:1997vv,Gava:1998sv}. Although the
supergravity picture breaks down in this limit one can still hope to
match some protected quantities like the BPS spectrum of states, two-
and three-point functions, etc. in the supergravity side with those in
the SCFT. We will give some evidence that this is indeed the case.

In the past, Kaluza-Klein spectra of supergravities have mainly been
studied for reductions on factorized geometries. Together
with~\cite{tseytlin}, the reductions on warped $AdS_3\times S^7$
presented here, to the best of our knowledge constitute the only
examples in which the full KK spectrum has been worked out. We hope
that the techniques developed here can help to improve this situation.
As we will see, the familiar relations between masses and dimensions
in pure $AdS$ spaces get modified by the presence of warped factors
and non-trivial dilatons.  We work out the details of these
modifications for the general case of domain wall/QFT correspondences
between supergravity on warped $AdS_{d+1}\times S^q$ and SYM theories
with $16$ supercharges.

The paper is organized as follows: In section~2, we extend the
familiar holographic relations between masses and conformal dimensions
to the case of supergravities on warped $AdS_{d+1}\times S^q$. We
present a sample calculation of the scalar two-point function in the
warped $AdS$ and discuss the conformal isometries of the warped string
background.  Sections~3, 4, and 5 are devoted to the study of KK
harmonics of type IIA, IIB and type I supergravity, respectively, on
warped $AdS_3\times S^7$. We start by deriving the spectrum of sphere
harmonics by a simple group theory analysis and confirm the full
supermultiplet structure by explicit calculation of the linearized
field equations.
In Sections~6 we compare the supergravity results with the
expectations from matrix string theory.  In section~7 we comment on
interesting directions of future research.  The appendix combines a
series of tables displaying the quantum numbers and structure of
supermultiplets relevant for the discussions in the main text.

\mathon
\section{Warped $AdS_{d+1}\times S^q$  geometries}
\mathoff

In this section we apply the ideas of \cite{Witten:1998qj} to extend
the dictionary between masses and scaling dimensions on $AdS$ spaces
to domain wall/QFT correspondences involving warped $AdS_{d+1}\times
S^q$ geometries. More specifically, we consider spaces described by a
metric
\bea
d\hat{s}^2&=& z^\omega \, ds^2=z^\omega \,\left[ {\ell^2\over z^2}\,
dx^\mu\, dx_\mu+\tilde{\ell}^2\, d\Omega_q\right]\;.\label{metricp}
\eea
Greek letters $\mu=0, 1, \dots, d$, refer to components along
$AdS_{d+1}$ with $x^{d}\equiv z$, while Arabic indices $m=d+1,
\dots, D-1$ run over the $S^q$ sphere with metric element
$d\Omega_q$. Geometries of the general form (\ref{metricp}) typically
arise as near horizon geometries of p-brane solutions of low energy
supergravities. In addition, they involve a non-trivial flux for a
rank $d\pls1$ form and --- unless $(D,p)= (10,3), (11,2), (11,5)$ ---
a running dilaton.  For definiteness, let us consider Dp-branes, such
that $D=10$, $d=p+1$, $q=9-d$, and (for $d\neq 6$)
\bea
\omega &=&-{(d-4)^2\over 4\,(6-d)} \;, \qquad
\tilde{\ell}^2~=~{\ell^2\over 4}(6-d)^2 \;, \qquad
z =~ ~{2\sqrt{c_d\, g_{\rm YM}^2\, N}\over 6-d}\, r^{d-6\over 2}
\;,
\label{wDp}
\eea
\\
with a constant $c_d$, and $r$ denoting the distance from the brane
source. The Yang-Mills coupling constant $g_{\rm YM}$ is given by
$g_{\rm YM}^2=2(2\pi)^{d-3} g_s (\alpha^\prime)^{d-4\over 2}$ and
carries dimension of $[L]^{d-4}$. Perturbation theory is better
organized in terms the dimensionless 't Hooft parameter
\bea
\lambda_{\rm eff}&\equiv& c_d\,g_{\rm YM}^2\, N\, \left({r\over
\alpha^\prime}\right)^{d-4} \;.
\eea
The gauge theory is weakly coupled for $\lambda_{\rm eff}\ll 1$.
The other two relevant parameters are
\be \ell^2\, z^\omega\sim\sqrt{N}\, \lambda_{\rm
eff}^{d-4\over4}\;, \qquad e^{\phi}\sim{\lambda_{\rm
eff}^{8-d\over 4}\over N} \;, \label{lw} \ee
denoting the $AdS$ radius and the string coupling constant,
cf.~\cite{Itzhaki:1998dd}.  As usual, genus expansion corresponds to
an expansion in ${1\over N}$. We will always work in the limit of
large $N$ with $\lambda_{\rm eff}$ kept fixed where both supergravity
and string perturbation can be trusted.

\subsection{Scalar fields}

In this and the next subsection we derive the relations between
masses and scaling dimensions for fields moving on warped $AdS$
geometries. For similar results on pure $AdS$ spaces see
\cite{Witten:1998qj,Henningson:1998cd}.
We start by considering a massless scalar field moving freely on
(\ref{metricp}). Rewriting the warped d'Alambertian in terms of the
pure $AdS_{d+1}$ one, the scalar equation of motion takes the form
\bea
 z^{\omega}\,\hat{\square}_D\, \phi &\equiv &
{z^{\omega}\over \hat{e}}\,\partial_M\,
\left(\hat{e}\,\hat{g}^{MN}\,\partial_N\,\phi
\right)\nn\\
&=& \left(\square_{AdS}+{\omega\over 2\,
\ell^2}\,(D-2)\,\partial_z - m^2\right)\, \phi=0 \;.
\label{scalar} \eea
Here and in the following we denote by hats those quantities computed
using the warped metric $\hat{g}_{MN}=z^\omega \,g_{MN}$
(\ref{metricp}). The mass parameter $m$ is defined by the eigenvalue
equation:
\bea
\square_{S^q}\,\phi=-m^2\, \phi \;,
\nn
\eea
and characterizes the harmonic mode of the scalar field $\phi$ along
the sphere $S^q$. After a Fourier transform in the $d$-dimensional
space spanned by ${\bf x}=(x^0, \dots, x^{d-1})$
\be
\phi(x^{\mu})=\int \, d{\bf p}\,e^{i {\bf p}\,{\bf x}}\,\phi_p(z)
\;,
\la{Four}
\ee
the scalar equation (\ref{scalar}) reduces to a second order
differential equation for $\phi_p(z)$ of the kind
\bea
\left(z^2\, \dd_z^2 + \left(1-2\,a \right)\, z\,\dd_z +
p^2\,z^2+a^2-\delta^2 \right)\,\phi_p(z) \;,
\label{diffeq}
\eea
with
\bea
 a ~=~{d\over 2}-{\omega\over 4}\,(D-2)
\;,\qquad
\delta ~=~ \sqrt{a^2+ m^2 \ell^2} \;.
\label{cs}
\eea
Its general solution can be expressed in terms of Bessel functions
$J_\delta, Y_\delta$, as
\bea \phi_p(z) &=& z^{a}\left[ d_1 \, J_\delta (pz) +
d_2\,Y_\delta(pz)\right]\;. \label{Bessel} \eea

The ``conformal'' dimension $\Delta$ of the holographically related
operator ${\cal O}_\Delta$ can be extracted from the scaling behavior
of the two solutions \Ref{Bessel} for $\phi_p(z)$ at the boundary
$z\sim 0$ as
\bea
\phi_p &\sim& \left\{ \begin{array}{r}
z^{E_0-2\,\delta} \left( \phi_{\rm def} + \CO(z) \right) \\
z^{E_0} \left( \phi_{\rm vev} + \CO(z) \right)
\end{array}\right. \;,
\la{scalz}
\eea
with
\bea E_0&=& a+\delta=a+\sqrt{a^2+m^2\, \ell^2} \;, \nn\\[1ex]
\Delta&=& {E_0\over 2}\,(6-d) \;. \label{u0} \eea
The last equation relates the quantum number $E_0$ which measures the
power behavior in $z\sim r^{d-6\over 2}$ of $\phi$ near the boundary
to the conformal dimension $\Delta$ associated to rescalings of the
brane distance $r$.  In the conformal case $\omega=0$, $d=4$, the two
quantities coincide and we recover the familiar relation between
scalar masses and conformal dimensions in $AdS_5$ with
$a\equ{d\over2}\equ2$. In a general near horizon Dp-brane geometry one
finds instead
\bea
a
&=&{8-d\over 6-d}\;,\qquad
\Delta ~=~ {\tilde{d}\over 2}+{1\over 2}\sqrt{\tilde{d}^2+ 4\,
m^2\,\tilde{\ell}^2}
\label{md0}
\eea
where $\tilde{d}=8-d$ denotes the worldvolume dimension of the
magnetically dual brane. Note that for $d=2$ and $d=5$, these
relations imply $a ={d+1\over 2}$, leading to a differential equation
\Ref{diffeq} which coincides with that for pure $AdS$ space in
dimension $d\pls1$. This reflects the fact that the near horizon
geometries of fundamental strings and D4-branes of type IIA theory can
be derived via dimensional reduction from those of M2 and M5 branes,
respectively, a fact that will be extensively exploited later.

It is important to notice that the differential equation
(\ref{diffeq}) keeps its form under rescaling of the field $\phi_p(z)$
\bea
\phi_p(z) &\rightarrow& z^\xi\phi_p(z)\;,
\label{xi}
\eea
while the parameters get modified as $a\rightarrow a-\xi$,
$\delta\rightarrow \delta$\,.  Likewise, we note that the form
of~(\ref{diffeq}) remains invariant under
\bea \phi_p(z) &\rightarrow& \phi_p(z) -
\frac{z}{a\pm\delta}\,\phi_p'(z) \;, \eea
under which its parameters transform as $a\rightarrow a-1$,
$\delta\rightarrow \delta\mp 1$\,.

\subsection{Higher spin modes}

Following \cite{Witten:1998qj,Henningson:1998cd} one can easily extend
the above dictionary between masses and conformal dimensions to higher
spin fields. One starts by constructing Green functions $\Phi_0(z)$
defined as a solution of the field equations depending only on $z$
(besides possible non trivial harmonic dependence on the sphere
coordinates $y$ which will be implicitly understood). The dependence
on ${\bf x}$ can be restored later on by means of an $SO(1,d+1)$
transformation $x_\mu \rightarrow {x_\mu \over x^2}$. Here we restrict
ourselves to determined the $z$-dependence of the Green function
associated i.e.\ the mode ${\bf p}=0$.

For massless spin ${1\over 2}$ fields $\hat{\Psi}_0(z)$ in $D$
dimensions one finds
\bea z^{\omega\over
2}\,\hat{\Gamma}^M\,\hat{D}_M\,\hat{\Psi}_0&=&
\left(\Gamma^\mu\,D_\mu+\Gamma^m\,D_m+{\omega\over
4\, z}\,(D-1)\, \Gamma^z \right)\hat{\Psi}_0\nn\\
&=&
z^{-1}\,\Gamma^z\,\left(z\,\partial_z+(a-{\omega\over 4})\,\pm
m\ell \right)\hat{\Psi}_0\nn\\
&=& z^{-1+{\omega\over 4}}\,\Gamma^z\,\left(\partial_z+a\,\pm
m\ell\right)\Psi_0~=~ 0 \;,
\label{s1/2}
\eea
where in the last line we have introduced the rescaled field
$\Phi\equiv z^{\omega\over 4}\, \hat{\Psi}$ in such a way that the
value of $a$ in (\ref{s1/2}) matches the one found for scalar fields
in (\ref{cs}). The mass parameter $m$ is again defined in terms of the
sphere harmonics as
\ben
\Gamma^m D_m\, \Psi_0=\pm m\,\Psi_0 \;,
\een
with $SO(7)$ gamma matrices $\Gamma^m$.  The scaling of the two
solutions of (\ref{s1/2}) near the boundary $z\sim 0$ are
\be \Psi_0(z)\sim z^{a\pm m\ell} \;.
\ee
Using (\ref{scalz}), we can identify $E_0-2\delta$ and $E_0$ as the
powers of $z$ from which we read off the mass/dimension relation
\bea
E_0&=&a+|m\, \ell|\;,\qquad
a ~=~{d\over 2}-{\omega\over 4}\,(D-2) \;.
\label{md12}
\eea
A similar analysis applies to higher spin modes. As an illustration of
these cases let us consider a rank $n$ form
\bea
\hat{A}_{\mu_1\dots\mu_p m_{p+1}\dots m_n}(z)
&=&e_{\mu_1}^{\alpha_1}\dots
e_{m_{p+1}}^{a_{p+1}}\,A_{\alpha_1\dots\alpha_p a_{p+1}\dots a_n}(z)
\eea
with $p$ legs along $AdS_{d+1}$. It is important to notice that unlike
in more familiar $AdS$/CFT instances the scaling dimension of the dual
operators associated to $p$ forms in the warped $AdS_{d+1}$ with
different higher dimensional origins, are different due to extra
$\omega$-shifts coming from every leg aligned along the warped
sphere. For simplicity, we restrict ourselves to the case where only
one set of components $[\alpha_1\dots\alpha_p a_{p+1}\dots a_n]$ with
$\alpha_i\neq z$ is turned on. To derive the asymptotic behavior, we
start from the ansatz
\be
A_{\alpha_1\dots\alpha_p a_{p+1}\dots a_n}(z)=c\,
z^\lambda\,\epsilon_{\alpha_1\dots\alpha_p a_{p+1}\dots a_n}
\;.
\ee
The equations of motion
\bea
\partial_{M_0}\,\left(\hat{e} \,\hat{g}^{M_0
N_0}\dots\hat{g}^{M_{n}N_n}\,\partial_{[N_0}\,
\hat{A}_{N_1\dots N_n]}\right) &=& 0 \;,
\eea
then reduce to a quadratic equation in $\lambda$ with solutions
$\lambda_+=E_0$, $\lambda_-=E_0-2\, \delta$ and
\bea
E_0&=&a+\delta=a+\sqrt{\left(a-p+{n\,\omega\over
2}\right)^2+m^2\,\ell^2}\;, \nn\\
 a&=&{d\over
2}-{\omega\over 4}\,(D-2) \;, \label{md1}
\eea
where the mass parameter $m$ is defined by the harmonic equation
\ben
D_m\, H^{m M_2\dots M_n}= -m^2\,A^{M_2\dots M_n} \;.
\een

The relations \Ref{md0}, \Ref{md12}, \Ref{md1} generalize the familiar
relations \cite{Henningson:1998cd} between masses of fields on $AdS$
and scaling dimensions of the dual operators to domain wall/QFT
involving warped geometries.  In the next sections we apply this
dictionary to the study of KK reductions of ten-dimensional strings on
near horizon string-like backgrounds.

\mathon
\subsection{Scalar two-point function}
\mathoff

The meaning of the quantum number $E_0$ (to which we will refer as the
energy or scaling dimension) in the non-conformal instances that we
study here, at first sight may seem obscure. As we mentioned in the
introduction (see also section 2.4 below), the $AdS$ group acts as the
group of conformal isometries of the domain wall background and
therefore gauge/supergravity quantities must transform covariantly
under conformal rescalings. It is natural to assign to supergravity
states the quantum number $E_0$ to describe these scaling properties.
Here we present a more technical definition of this quantity which
makes contact with the definition of conformal dimensions in more
familiar $AdS$/CFT instances. More precisely, we show that, under
suitable normalizations, the operator two-point function derived from
supergravity scalar fields on warped $AdS$ behaves like $\langle {\cal
O}({\bf x}){\cal O}({\bf y})\rangle\sim |{\bf x}-{\bf y}|^{-2 E_0}$.
This is the expected result for a two-point function involving
operators of dimension $E_0$ in a dual scale invariant boundary
theory.

The derivation of the scalar two-point function in warped $AdS$ is a
straightforward generalization of the similar calculation in pure
$AdS$ (see for example the Appendix of \cite{Freedman:1998tz}). The
action for a free massive scalar on warped $AdS_{d+1}$ reads
\bea S&=&\ft12\, \int
d^{d+1}x\,\hat{e}
\left(\hat{g}^{\mu\nu}\,\partial_{\mu}\phi\,\partial_{\nu}
\phi+m^2 z^{-\omega} \phi^2\right)\nn\\
&=&\ft12 \,\int d^{d+1}x\,e\,z^{\omega(d-1)\over 2}\, \left(
g^{\mu\nu}\,\partial_{\mu}\phi\,\partial_{\nu}\phi+  m^2
\phi^2\right) \;.
\label{massive}
\eea
The equations of motion coming from (\ref{massive}) were solved in the
previous section.  After Wick rotation and choosing appropriate
boundary conditions they can be written as
\be \phi(x^{\mu})=\int \, d{\bf p}\,e^{i {\bf
p}\,{\bf x}}\,\phi_p(z)= \int \, d{\bf p}\,e^{i {\bf p}\,{\bf
x}}\,\phi_B(p)\,K_\delta^\epsilon(z|p) \;,
\label{ans2}
\ee
with the renormalized Bessel function
\ben
K_\delta^\epsilon(z|p)\equiv \left( {z\over \epsilon}\right)^a\,
{K_\delta(z|p)\over K_\delta(\epsilon|p)} \;,
\een
satisfying
\ben
\lim_{z\rightarrow \epsilon} K^\epsilon_\delta(z|p)=1\;,
 \;\;
\lim_{z\rightarrow \infty} K^\epsilon_\delta(z|p)=0  \;.
\een
Plugging (\ref{ans2}) into the action one finds after integration by
parts:
\ben
S=\ft12\int \, d{\bf p}d{\bf p'}\,\delta({\bf p}+{\bf
p'})\,\phi_B({\bf p})\phi_B({\bf
p'})\,\epsilon^{-2a+1}\,\lim_{z\rightarrow
\epsilon}\,\partial_z\,K_\delta^\epsilon(z|p) \;,
\een
from which one reads the two-point function (see \cite{Freedman:1998tz}
for details):
\bea
\langle {\cal O}({\bf p}){\cal O}({\bf
 p'})\rangle&=&\delta({\bf
p}+{\bf
p'})\,p^{2\delta}\,
\epsilon^{2(\delta-a)}\,2^{-2\delta}(-2\delta){\Gamma(1-\delta)\over
\Gamma(1+\delta)} \;.
\eea
After Fourier transform one is finally left with
\be \langle {\cal O}({\bf x}){\cal O}({\bf
 y})\rangle\sim |{\bf x}-{\bf y}|^{-2\left(\delta+\ft12 d\right)}
\;.
\label{cd}
\ee
Notice that the dimension read off from (\ref{cd}) matches $E_0$ in
(\ref{u0}) after a suitable rescaling of $\phi_p(z)$. As one can see
from (\ref{xi}), such rescalings shift $E_0$ while keeping the
argument under the square root in $\delta$ invariant. The whole
information about the warped geometry is contained in this
$\delta$. In particular, the mass bound at which the argument of the
square root becomes negative is shifted in the warped geometry with
respect to its pure $AdS$ cousin. It would be nice to supply this
observation with a stability analysis in the spirit
of~\cite{Breitenlohner:jf}. A rather non-trivial consistency check of
the whole picture follows from the need of a certain conspiracy
between masses and warped factors entering $\delta$ in (\ref{u0}) in
order to produce rational numbers as output of the square root. We
will verify this by explicit calculations of the spectrum of harmonics
in warped $AdS_3\times S^7$. In~\cite{Gherghetta:2001iv}, two-point
correlation functions have been determined by a similar computation
for the $SO(8)$ current and the stress-energy tensor, leading to
$\delta=\ft12$, and $\delta=\ft32$, respectively. As we will see below
--- cf.~\Ref{Evec15} and the subsequent discussion --- these values
are recovered in our general analysis; current and stress-energy
tensor couple to the $n=0$ components of spin $s_0=1, 2$ in table~5 of
the appendix.

It is worth to spend some words on our choice of normalizations.  As
we have seen in our sample calculation, an overall shift in the
definition of $E_0$ can be adjusted by a choice of frame. This is
clear from the fact that since the dilaton background carries an
explicit $z$-dependence, fields which are redefined by a
dilaton-dependent rescaling couple to operators of different
dimensions. In particular, fluctuations of the metric in the Einstein
or string frame carry different energies $E_0$. There is no a priori
privileged choice. We choose normalizations in such a way that all
field equations reduce to scalar type of equations with $E_0=a+\delta$
and $a$ given by (\ref{cs}). In the warped $AdS_3$ case this
correspond to choose $a=\ft32$. This choice is natural for the type
IIA case where string solutions descending from M2 branes by
dimensional reduction are naturally associated to a $d=3$ SCFT. For
the sake of comparison we adopt the same normalization in the type IIB
and type I case where no analog of the higher dimensional pure $AdS$
origin is available.

\mathon
\subsection{$AdS_3\times S^7$ vacua and conformal Killing isometries}
\mathoff

We will focus on string like solutions with near horizon geometries
conformal to $AdS_3\times S^7$, (for earlier investigations of these
backgrounds, see e.g.~\cite{Duff:fg}). These vacuum configurations
involve a non-trivial metric $g_{MN}$ , the dilaton $\phi$ and a rank
three form $H_3$ given by (here and in the following we always work in
the Einstein frame):
\bea
d\hat{s}^2&=& z^\omega \, ds^2=z^\omega \,\left[ {\ell^2\over
z^2}\,(-2\,dx_+\, dx_- +dz^2)+4\,\ell^2\, d\Omega_7\right]
\nn\\
e^{\Ge\phi}&=& \Phi_0\, z^{3\over 2}\nn\\
H_{01z}&=& {3\,\ell^2\over z^4}\label{vacuum}
\eea
with dimensionful constant $\Phi_0$, a sign $\Ge=-1,1$ for the
fundamental and D-string, respectively, and $\omega=-1/4$. The
variable $z$ is related to the D-brane distance $r$ via $z=r^{-2}$ and
therefore conformal dimensions $\Delta$ are given in terms of the
$E_0$ scaling dimension defined through (\ref{u0}) by
\be
\Delta=2\, E_0 \;.
\ee
The string vacuum preserves sixteen supercharges.  The number of
supersymmetries preserved by the near horizon geometry can be easily
understood by noticing that (\ref{vacuum}) arises from dimensional
reduction along $AdS_4$ of the maximal supersymmetric $AdS_4\times
S^7$ 11d-vacuum.  According to \cite{Lu:1996rh}, half of the $AdS_4$
Killing spinors are preserved by such reductions.

The geometry (\ref{vacuum}) is also invariant under $ISO(1,1)\times
SO(8)$ with $SO(8)$ the isometries of the seven sphere and $ISO(1,1)$
the two-dimensional Poincar\'e group.  In addition one can verify the
invariance under the following rescaling
\bea
x_{\mu}&\rightarrow& \lambda^2 \, x_{\mu}\;,\qquad
\ell^2 ~\rightarrow~ \lambda^{-2\,\omega} \,\ell^2\;.
\label{rescaling}
\eea
This is clearly not a symmetry of the theory since it relates the physics
at a given $AdS$ radius $\ell^2$ to that at $\lambda^{-2\,\omega}
\,\ell^2$. The existence of this invariance however requires that
quantities in the dual QFT should transform covariantly under
conformal rescalings. It is interesting to notice that the rescalings
(\ref{rescaling}) leave invariant the effective 't Hooft parameter
$\lambda_{\rm eff}$.

One can still go further and check that the full two-dimensional
conformal algebra arises as asymptotic conformal Killing isometries of
the warped geometry (\ref{vacuum}). The algebra is realized near the
boundary $z\sim 0$ (up to $\CO(z^4)$ terms) in terms of the following
two sets of Virasoro generators
\bea L_n
&=& x_+^{n+1}\,
\partial_+ +\ft14 z^2\,\ell^2\, x_+^{n-1}\,n(n+1)\,\partial_-
+\ft12 (n+1)\, x_+^n\,z\,\partial_z \;, \nn\\
\bar{L}_n &=& x_-^{n+1}\,
\partial_- +\ft14 z^2\,\ell^2\, x_-^{n-1}\,n(n+1)\,\partial_+
+\ft12(n+1)\, x_-^n\,z\,\partial_z \;,
\eea
which satisfy the conformal Killing equations
\be
\nabla_{(\mu}\, \xi_{\nu)}-{1\over 2}\,g_{\mu\nu}
\nabla^\rho\,\xi_\rho=\CO(z^\omega) \;,
\ee
up to terms of order $\CO(z^\omega)$ which at $z\sim 0$ fall off much
faster than the metric. Among these generators one can easily identify
the $ISO(1,1)$ Killing algebra with spin and translations realized by
$s_0=L_0-\bar{L}_0$, and $L_{-1},\bar{L}_{-1}$, respectively.
Finally, the scaling properties of fields/operators will be described
by $E_0=L_0+\bar{L}_0=z\partial_z$, which is clearly not a Killing but
a conformal Killing vector of (\ref{vacuum}).

The realization of the full two-dimensional conformal algebra near the
boundary $z\sim 0$ suggests that this string background represents
some sort of non-conformal deformation of a more fundamental dual SCFT
living at the $AdS_3$ boundary. The natural candidates for such dual
theories in the case of D-strings are the orbifold SCFT's
\cite{Dijkgraaf:1997vv,Gava:1998sv} to which the two-dimensional gauge
theories governing the dynamics of D-strings in type IIB and type I
theory flow in the infrared. They can be thought as second quantized
type II or heterotic strings moving on $(\bbR^8)^N/S_N$. Some striking
evidence for such a correspondence can be already be observed from the
scaling behavior of the dilaton in the string vacuum (\ref{vacuum})
\bea
e^{\phi} &=&  \Phi_0\, z^{3\over 2}\;,
\eea
implying that the associated operator is a scalar of conformal
dimension $\Delta= 2E_0= 3$. This suggests to identify this operator
with the one responsible for bringing the SCFT away from the conformal
point and identified in~\cite{Dijkgraaf:1997vv} as the $\bbZ_2$ twist
field with dimensions $(h,\bar{h})=(\ft32,\ft32)$, i.e.\ $s_0=0$ and
$\Delta=3$.  In this picture, the domain wall solution can be seen as
a deformation via the expectation value $g_s$ of a somewhat mysterious
conformal point.

\mathon
\section{Type IIA supergravity on warped $AdS_3 \times S^7$}
\mathoff

In this section, we compute the Kaluza-Klein spectrum of type IIA
supergravity on the warped $AdS_3 \times S^7$ background. We first
perform a simple group theory analysis to determine the spectrum of
sphere harmonics and applies to the type IIB case as well. To further
obtain the masses of the fields in the sense discussed in the last
section, we employ the $AdS_4 \times S^7$ vacuum of eleven-dimensional
supergravity whose spectrum has extensively been studied in the
eighties~\cite{Biran:1983iy}, and from which the warped $AdS_3 \times
S^7$ spectrum of type IIA descends upon further dimensional reduction.

\subsection{Group theory and sphere harmonics}

The spectrum of $SO(q+1)$ representations appearing in the
Kaluza-Klein reduction of a D-dimensional supergravity on the sphere
$S^q$ is essentially determined by group theory \cite{Salam:1981xd}
(see \cite{deboer} for a recent discussion in a similar context). The
harmonic expansion of a field transforming in the ${\cal R}_{SO(q)}$
representation of the Lorentz group of $S^q$ comprises all the
representations ${\cal R}_{SO(q+1)}$ of the isometry group $SO(q+1)$
that contain ${\cal R}_{SO(q)}$ in their decomposition. The presence
of a warp factor and a running dilaton does not alter this
analysis. In this section we apply this group theory analysis to
derive the spectrum of $SO(8)$ representations in the type IIA string
background.

\begin{table}[tb]
\centering
\begin{tabular}{c|ll}
\hline
$SO(7)$ & in $SO(8)$ & \\
\hline\hline \footnotesize 1
 & \footnotesize
$(n,0,0,0)$  & \footnotesize $n=0, 1, \dots$
\\ \hline
\footnotesize7
&\footnotesize       $(n,0,0,0)$ &
\footnotesize $n=1, 2, \dots$\\
&
\footnotesize$(n,1,0,0)$& \footnotesize  $n=0, 1, \dots$
\\ \hline
\footnotesize 21
&\footnotesize    $(n,0,1,1)$  &
\footnotesize  $n=0, 1, \dots$\\
&
\footnotesize   $(n,1,0,0)$  & \footnotesize  $n=0, 1, \dots$
\\ \hline
\footnotesize 27
&
\footnotesize    $(n,0,0,0)$  &
\footnotesize   $n=2, 3, \dots$ \\
&
\footnotesize    $(n,1,0,0)$  &
\footnotesize  $n=1, 2, \dots$ \\
&
\footnotesize  $(n,2,0,0)$  & \footnotesize   $n=0, 1, \dots$
\\ \hline
\footnotesize 8
& \footnotesize $(n,0,1,0)+ (n,0,0,1)$  & \footnotesize   $n=0, 1,
\dots$
\\ \hline
\footnotesize 35
& \footnotesize  $(n,0,2,0)+ (n,0,0,2)$  &
\footnotesize  $n=0, 1, \dots$ \\
               &
\footnotesize $(n,0,1,1)$  &
\footnotesize   $n=0, 1, \dots$
\\ \hline
\footnotesize 48
& \footnotesize   $(n,0,1,0) +(n,0,0,1)$ &
\footnotesize    $n=1, 2, \dots$ \\
&
\footnotesize  $(n,1,1,0)+ (n,1,0,1)$ &
\footnotesize    $n=0, 1,
\dots$
\\ \hline
\end{tabular}
\caption{\small Embedding of $SO(7)$ representations into $SO(8)$. }
\label{SO78}
\end{table}

The on-shell field content of ten-dimensional type IIA supergravity
$({\bf 8_v}\pls{\bf 8_s})\times({\bf 8_v}\pls{\bf 8_c})$ decomposes
under the $SO(7)$ Lorentz group of $S^7$ as
\bea
{\cal R}_{SO(7)}^{\rm II} &=&
3\cdot {\bf 1}+3\cdot {\bf 7}+4\cdot {\bf 8}+
2\cdot {\bf 21}+{\bf 27}+{\bf 35}+2\cdot {\bf 48} \;.
\label{so7a}
\eea
Every representation in (\ref{so7a}) gives rise to a tower of
harmonics of those $SO(8)$ representations in which it is contained
upon decomposition under $SO(7)$. We have summarized these series in
table~\ref{SO78}.
Together, this gives the following $SO(8)$ field content of type IIA
supergravity on $AdS_3\times S^7$:
\bea
\begin{tabular}{l|c|c|c}
\hline
$SO(8)$ &     $n=0$    &    $n=1$  & $n>1$ \\ \hline
$(0,0,0,n)$  &   3     &     6 & 7 \\
$(0,1,0,n)$  &    5      &    6 & 6\\
$(1,0,1,n)$  &   3      &    3 & 3\\
$(0,0,1,n)$  &    4      &    6 & 6\\
$(1,0,0,n)$  &    4      &    6 & 6 \\
$(0,2,0,n)$  &   1      &    1 & 1 \\
$(0,0,2,n)$  &   1      &    1 & 1\\
$(2,0,0,n)$  &   1     &     1 & 1\\
$(0,1,1,n)$  &   2     &     2 & 2 \\
$(1,1,0,n)$  &   2     &     2 & 2 \\
\hline
\end{tabular}
\la{specGT}
\eea
This spectrum may be conveniently organized in terms of
supermultiplets of the superalgebra of background isometries, which as
discussed above is build from the semidirect product of
$ISO(1,1)\times SO(8)$ and $16$ real supercharges. Moreover, we may
even assign to all the fields quantum numbers of the conformal
isometry group $SL(2,\bbR)\times OSp(8|2,\bbR)$. To analyze this
multiplet structure it turns out to be useful to employ the
representation structure of the well known spectrum of
eleven-dimensional supergravity on $S^7$.

\mathon
\subsection{Linearized equation of motions via reduction from $AdS_4$}
\mathoff

The warped $AdS_3\times S^7$ string solution of type IIA supergravity
descends from the much better studied $AdS_4\times S^7$ vacuum of
eleven-dimensional supergravity upon dimensional reduction along one of
the $AdS_4$ coordinates. The spectrum of KK harmonics and linearized
equations of motion around the type IIA string background can then
easily be derived from the $AdS_4$ results \cite{Biran:1983iy} upon a
further dimensional reduction. We focus on bosonic fluctuations. Field
equations for scalar, vector and spin-2 fields in $AdS_4$ take the
form
\bea
\square\, \phi &=&
\ft{1}{\tilde{\ell}^2}(\lambda-8)\, \phi \;, \la{4dsca}\\
D^M F_{MN} &=& \ft{1}{\tilde{\ell}^2}\lambda\,A_N \;, \la{4dvec} \\
\square\, \eta_{MN}&=& \ft{1}{\tilde{\ell}^2}(\lambda-16)\,\eta_{MN}
\;, \qquad (\eta^M{}_M = 0\;,\quad D^M\eta_{MN}=0) \;.
\la{4dsp2}
\eea
with capital indices $M=\mu,y$, running over the 4d spacetime, and
$\mu=0,1,2$. The spectrum of masses $\lambda$ has been derived
in~\cite{Biran:1983iy}. It is organized in supermultiplets of the
group $OSp(8|4,\bbR)$. Upon dimensional reduction in $y$, the scalar
equation \Ref{4dsca} reduces to
\bea
\square_{AdS_3} \left( z^{-\frac12} \, \phi \right) &=&
\ft{1}{\tilde{\ell}^2}(\lambda\mis3)\, z^{-\frac12}\,\phi \;.
\label{34sca}
\eea
Evaluating this equation with the ansatz \Ref{Four}
$\phi(x^{\mu})=e^{i {\bf p}\,{\bf x}}\,\phi_p(z)$, this leads to a
three-dimensional equation of the type \Ref{diffeq}, with parameters
\bea
a &=& \ft32\;,\qquad \delta~=~\ft12{\sqrt{1+\lambda}} \;,\qquad
E_0=\ft32+\ft12\sqrt{1+\lambda} \;,
\label{cha}
\eea
whose solution may be given in terms of Bessel functions. Notice that
according to our discussion above, a differential equation with these
characteristics coincides with the one coming from KK reduction of
massless fields on warped $AdS_3\times S^7$, cf.~(\ref{cs}) with
$\omega=-\ft14$, $D=10$. This is clear from the fact that these
results could be derived directly from reduction of type IIA on the
warped string background. In the following, we will choose~\Ref{34sca}
as the standard equation for scalar fields on the warped $AdS_3$
background.

The four-dimensional vector equation \Ref{4dvec} with the ansatz
$A_M = (A_\mu, e_y{}^y A_y)=(A_\mu, {\ell\over z}\,A_y)$ turns into
\bea
\square_{AdS_3} \Big( z^{-\frac12} \, A_y \Big) &=&
\ft{1}{\tilde{\ell}^2}(\lambda-3)\, z^{-\frac12}\,A_y \;, \non
D^\mu \left( z^{-1}\,F_{\mu\nu}\right) &=&
\ft{1}{\tilde{\ell}^2}\lambda\,z^{-1}\,A_\nu \;,
\label{3dvec}
\eea
on $AdS_3$. As expected, the component $A_y$ decouples and gives rise
to a scalar equation whose form coincides with \Ref{cha}. The
three-dimensional vector equation may again be solved in terms of
Bessel functions as
\bea
A_\mu &=& (A_+, A_-, A_z) ~=~
\left(
\ft{p_+}{z} (a_1 - \ft12 z a_1' + a_2) ,\,
\ft{p_-}{z} (a_1 - \ft12 z a_1' - a_2),\,
p_+p_- w \right) \;,
\eea
where $a_{1,2}$ are solutions of the scalar field equation
\Ref{diffeq} with parameters \Ref{cha}. The energy may be read off
from the scaling behavior of the flat components
$\tilde{A}_\pm=e_\pm{}^\mu\, A_\mu\sim z^{E_0}$ in the limit $z\sim
0$, cf.~\cite{Witten:1998qj}, and coincides with that of the scalar
component \Ref{3dvec}. As expected, a four-dimensional vector on $AdS$
upon dimensional reduction hence gives rise to a three-dimensional
scalar and the two degrees of freedom of a massive spin 1 field,
sharing the same energy $E_0$. Accordingly, we will assign to these
fields quantum numbers $(\ell_0,\bar{\ell}_0)$ of the
three-dimensional conformal isometry group $SO(2,2)$ as
\be
\left(\ft12E_0,\ft12E_0\right)\;,\quad
\left(\ft12E_0\mp\ft12,\ft12E_0\pm\ft12\right) \;,
\la{4dNvec}
\ee
with $E_0=\ft32+\ft12\sqrt{1+\lambda}\,$.
Similarly, the reduction of the spin-2 field equations \Ref{4dsp2}
yields a coupled system of a spin-0, spin-1 and spin-2 fields in
warped $AdS_3$. At ${\bf p}=0$ we find again scalar equations of the
standard form (\ref{cs}) with characteristics (\ref{cha}). I.e.\ this
field gives rise to three-dimensional fields with $SO(2,2)$ quantum
numbers
\be
\left(\ft12E_0,\ft12E_0\right)\;,\quad
\left(\ft12E_0\mp\ft12,\ft12E_0\pm\ft12\right) \;,\quad
\left(\ft12E_0\mp1,\ft12E_0\pm1\right) \;.
\la{4dNsp2}
\ee
Supplying the fields \Ref{specGT} with quantum numbers according to
\Ref{4dNvec}, \Ref{4dNsp2} suggests a grouping into the ${\cal
N}=(8,8)$ supermultiplets ${\bf (n000)}_{\rm IIA}$ defined in
table~\ref{specIIA}. The state in the upper left corner has
$(\ell_0,\bar{\ell}_0)=(\ft14(n\pls2),\ft14(n\pls2))$. The 8 unbroken
supersymmetry generators act vertically in this table, increasing the
value of $\bar{\ell}_0$ from top to bottom by $\ft12$ per row. The
value of $\ell_0$ is increased from left to right by $\ft12$ per
column which may be thought of as the action of the broken
supersymmetry generators. The omission of any representation with
negative Dynkin labels will be always understood. The full
spectrum~\Ref{specGT} may then be written as
\bea
{\cal H}_{\rm IIA}=
 \sum_{n=0}^{\infty} {\bf (n000)}_{\rm IIA}= \sum_{n=0}^{\infty}
 ({\bf 8_v} - {\bf 8_s})({\bf 8_v} - {\bf 8_s})(n000)
\label{sumA}
\eea
with products of $SO(8)$ representations always understood as tensor
products. The nongeneric multiplicities in \Ref{specGT} for small $n$
precisely match with \Ref{sumA}, taking into account that
four-dimensional massless vector and spin-2 fields give rise to only
two physical degrees of freedom in \Ref{4dNvec}, \Ref{4dNsp2} due to
the additional gauge freedom in \Ref{4dvec}, \Ref{4dsp2}. In
particular, for $n=0$ the physical degrees of freedom are entirely
contained in first two columns of table~\ref{specIIA}; $n=0$ states in
the remaining three columns correspond to pure gauge degrees of
freedom. The restriction to the first column at $n=0$ has been argued
to be a consistent truncation of supergravity on
$S^7$~\cite{Cvetic:2000dm}.

\begin{table}[bt]
\centering
\begin{tabular}{||l|l|l|l|l||}
\hline \hline \footnotesize $(n\plss2000)$ & \footnotesize
$(n\plss1010)$ & \footnotesize $(n100)$ & \footnotesize $(n001)$ &
\footnotesize $(n000)$ \\
\footnotesize $(n\pls1010)$ & \footnotesize $(n100)\pls(n020)$ &
\footnotesize $(n001)\pls(n\miss1 110)$ & \footnotesize
$(n000)\pls(n\miss1 011)$  &
\footnotesize $(n\miss1 010)$ \\
\footnotesize $(n100)$ & \footnotesize $(n001)\pls(n\miss1 110)$ &
\footnotesize $(n000)\pls(n\miss1 011)\pls(n\miss2 200)$ &
\footnotesize $(n\miss1 010)\pls(n\miss2 101)$  &
\footnotesize $(n\miss2 100)$ \\
\footnotesize $(n001)$ & \footnotesize $(n000)\pls(n\miss1 011)$ &
\footnotesize $(n\miss1 010)\pls(n\miss2 101)$ & \footnotesize
$(n\miss2 100)\pls(n\miss2 002)$  &
\footnotesize $(n\miss2 001)$ \\
\footnotesize $(n000)$ & \footnotesize $(n\miss1 010)$ &
\footnotesize $(n\miss2 100)$ & \footnotesize $(n\miss2 001)$  &
\footnotesize $(n\miss2 000)$ \\
\hline \hline
\end{tabular}
\caption{\small Spectrum of IIA supergravity on $S^7$, the multiplet
${\bf (n000)}_{\rm IIA}$.}
\label{specIIA}
\end{table}

Alternatively, these results can be derived by decomposing short
multiplets of the $AdS_4$ supergroup $OSp(8|4;R)$ \cite{Gunaydin:tc}
in terms of supermultiplets of its $SL(2,\bbR)_L\times
OSp(8|2;\bbR)_R$ subgroup \cite{Gunaydin:1986fe}, which is the
conformal isometry of the background. The $ISO(1,1)\times SO(8)$
isometry group is instead generated by $L_{-1}, \bar{L}_{-1},
s_0=L_0-\bar{L}_0$ and $S^7$ Killing isometries. As we explain in more
detail in the appendix, the columns of table~\ref{specIIA} in fact
correspond to primary subsets of the long multiplets of
$SL(2,\bbR)_L\times OSp(8|2;\bbR)_R$, cf.~tables~5--9. The $E_0$
quantum numbers refer to the eigenvalues of
$L_0+\bar{L}_0=z\,\partial_z$.

In order to facilitate the comparison with the spectra of the other
ten-dimensional supergravities, it is convenient to rewrite the
spectrum in terms of ${\cal N}=(8,0)$ supermultiplets defined through
\be
{\bf (n_1 n_2 n_3 n_4)_s}\equiv ({\bf 8_v} -
{\bf 8_s})(n_1 n_2 n_3 n_4) \;.
\la{multI}
\ee
In terms of these multiplets, the sum (\ref{sumA}) read
\be {\cal H}_{\rm IIA}=
\sum_{n=0}^{\infty}\left[{\bf (n+2000)_s}+{\bf (n+1001)_s}+
{\bf (n100)_s}+{\bf (n010)_s}+{\bf (n000)_s}\right]\;,
\ee
with every ${\cal N}=(8,0)$ multiplet spanning a column in
table~\ref{specIIA}.

\mathon
\section{Type IIB supergravity on warped $AdS_3 \times S^7$}
\mathoff

The KK harmonic analysis for type IIB on warped $AdS_3\times S^7$
closely follows the one for the type IIA case above with some
important differences. After reduction to $SO(7)$, the type IIB field
content $({\bf 8_v}+{\bf 8_s})\times({\bf 8_v}+{\bf 8_s})$ again gives
rise to (\ref{so7a}) and therefore to the same tower of $SO(8)$ states
\Ref{specGT}. The structure of the supermultiplets however is
substantially different. In contrast with the type IIA result
(\ref{sumA}), we expect for type IIB a decomposition according to
\bea
{\cal H}_{\rm IIB}= \sum_{n=0}^{\infty} {\bf (n000)}_{\rm
IIB}=\sum_{n=0}^{\infty} ({\bf 8_v} - {\bf 8_s})({\bf 8_v} -
{\bf 8_c})(n000) \;,
\label{sumB}
\eea
i.e.\ with a different assignment of quantum numbers
$(\ell_0,\bar{\ell}_0)$ according to the flip of chirality in
\Ref{sumB}. These numbers may be extracted from table~\ref{specIIB},
where again the value of $\bar{\ell}_0$ increases from top to bottom
by $\ft12$ per row whereas the value of $\ell_0$ is increased from
left to right by $\ft12$ per column. It is related to
table~\ref{specIIA} by interchanging the second and fourth column and
simultaneous shift of $n$ in these two columns.

Note that the two infinite sums (\ref{sumA}) and (\ref{sumB}) coincide
and agree with \Ref{specGT}, although they correspond to different
groupings of the fields into supermultiplets. The supermultiplets
${\bf (n000)}_{\rm IIA}$ and ${\bf (n000)}_{\rm IIB}$ share the first,
third and fifth columns in tables~\ref{specIIA}
and~\ref{specIIB}. Bosonic modes in these columns are associated to
the NSNS sector common to all five fundamental string theories. The
associated states can be directly read off from the type IIA results
in the previous section.

\begin{table}[bt]
\centering
\begin{tabular}{||l|l|l|l|l||}
\hline \hline \footnotesize $(n\plss2000)$ & \footnotesize
$(n\plss1001)$ & \footnotesize $(n100)$ & \footnotesize $(n010)$ &
\footnotesize $(n000)$ \\
\footnotesize $(n\pls1010)$ & \footnotesize
$(n\plss1000)\pls(n011)$ & \footnotesize $(n001)\pls(n\miss1 110)$
& \footnotesize $(n\miss1100)\pls(n\miss1 020)$  &
\footnotesize $(n\miss1 010)$ \\
\footnotesize $(n100)$ & \footnotesize $(n010)\pls(n\miss1 101)$ &
\footnotesize $(n000)\pls(n\miss1 011)\pls(n\miss2 200)$ &
\footnotesize $(n\miss1 001)\pls(n\miss2 110)$  &
\footnotesize $(n\miss2 100)$ \\
\footnotesize $(n001)$ & \footnotesize $(n\mis1100)\pls(n\miss1
002)$ & \footnotesize $(n\miss1 010)\pls(n\miss2 101)$ &
\footnotesize $(n\miss1000)\pls(n\miss2 011)$  &
\footnotesize $(n\miss2 001)$ \\
\footnotesize $(n000)$ & \footnotesize $(n\miss1 001)$ &
\footnotesize $(n\miss2 100)$ & \footnotesize $(n\miss2 010)$  &
\footnotesize $(n\miss2 000)$ \\
\hline \hline
\end{tabular}
\caption{\small Spectrum of IIB supergravity on $S^7$, the multiplet
${\bf (n000)}_{\rm IIB}$. }
\label{specIIB}
\end{table}

In the rest of this section, we shall confirm this structure of the
IIB spectrum, by explicitly computing the linearized field equations
around the type IIB string background. Similar KK techniques have been
applied in \cite{Deger:1998nm} to the study of supergravity harmonics
on $AdS_3\times S^3$. We will mainly concentrate on the states
corresponding to the second and fourth column of table~\ref{specIIB}
which cannot be derived from the IIA results of the last section,
cf.~appendix, tables~6,~8. For definiteness, we will specify the
analysis in this section to the D-string case. The spectrum associated
to the supergravity harmonics around the fundamental string background
is clearly the same, while NSNS and RR two-forms get exchanged under
S-duality.

\subsection{Equations of motion and background}

The bosonic field equations for type IIB supergravity can be
written as~\cite{Schwarz:qr}
\bea
D^MP_M &=& \ft1{24} G_{MNP}G^{MNP} \nn\\
D^PG_{MNP} &=& -P^P G^*_{MNP}-{2\over 3}\,i\,F_{MNPQR}\,G^{PQR}
\non
-R_{MN} &=& P_MP^*_N + P_M^*P_N +{1\over 6}\,F_{MP_1\dots
P_4}\,F^{P_1\dots P_4}_N\nn\\
&& \ft18 \left( G_M{}^{PQ}G^*_{NPQ} + G^*_M{}^{PQ}G_{NPQ} - \ft16
g_{MN} G^{PQR}G^*_{PQR} \right)\nn\\
F_{M_1\dots M_5}&=&{1\over 5!}\, \omega_{M_1\dots M_5
N_1\dots N_5}\,F^{N_1\dots N_5}
\;,
 \label{eqIIB}
\eea
with  field strengths
\bea
G_{MNP} &=& -\epsilon_{\alpha\beta}\,V^\alpha{}_+\, F_{MNP}^\beta
\;,
\nn\\
F_{MNPQR}&=&5\,\partial_{[M}\, A_{NPQR]}+
{5\over 8}\,i\epsilon_{\alpha\beta}\,
A^\alpha_{[MN}\,F^\beta_{PQR]} \;,
\eea
and $F^1=F^{2\,*}$. Here and in the following, we denote by
$\omega_{M_1\dots M_D}=e\,\epsilon_{M_1\dots M_D}$ the $D$-dimensional
volume form. The dilaton-axion system $\phi$, $C_0$ is encoded in the
$SU(1,1)$ matrix
\bea
V^\alpha{}_\pm &=& \frac1{2\sqrt{\tau_2}} \left(
\begin{array}{cc}
i\bar{\tau}-1 & i\tau-1 \\
i\bar{\tau}+1 & i\tau+1
\end{array}
\right) \;,\qquad\tau = \tau_1+i\, \tau_2=C_0 + i e^{-\phi}
\;,
\nn\\
P_M &=& \frac{i\dd_M\tau}{2\tau_2} \;,\quad Q_M ~=~
-\frac{\dd_M\tau_1}{2\tau_2} \;.
\eea
The D-string background \Ref{vacuum} in these variables is given by
\be
\mathring{\tau} = i z^{-\frac32}\;,\qquad
\mathring{P}{}_z = \ft34z^{-1}\;,\qquad
\mathring{G}_{+-z} = 3\, z^{-\ft{13}{4}}\,\ell^2 \;,
\la{Dback}
\ee
and metric \Ref{vacuum}. It preserves half of the original
supersymmetry. We will expand the ten-dimensional fluctuations in
terms of sphere harmonics $Y_{\bf m}^{\ell}(y)$
\be
\Phi_{\bf \mu m}=\sum_{\ell}
\phi_\mu(x)\,Y_{\bf m}^{\ell}(y)
\ee
with collective indices ${\bf \mu}, {\bf m}$ carrying the $SO(3)\times
SO(7)$ Lorentz representation of a given field. The sum over ${\ell}$
run over all possible $SO(8)$ representations entering in the harmonic
decomposition, cf.\ table~\ref{SO78} and will typically be described
by an integer $k$. For clearness, we will in general suppress the
indication of the explicit dependence of $\phi(x)$ on $x^\mu$ and
$Y_{\bf m}^{\ell}(y)$ on $y^m$, as well as the index $\ell$, whenever
non-ambiguous.

\subsection{Three-dimensional equations}

After reduction on $S^7$, the bosonic ten-dimensional equations
(\ref{eqIIB}) give rises to a set of equations associated to particles
of spin 0, 1, 2 moving on warped $AdS_3$. The arising scalar field
equations will be of the general form \Ref{diffeq}. As we have
discussed above, this form of equation is compatible with a rescaling
\Ref{xi} under which the energy of the field changes --- in agreement
with the fact that the energy is associated with $L_0+\bar{L}_0 =
z\partial_z$ which is not a Killing vector of the background. In order
to allow a comparison between the different scalar equations, we will
transform all of them into the form of the modified second order
equation on pure $AdS$ space \Ref{34sca}, which naturally appeared in
the type IIA reduction, cf.\ also \Ref{diffeq}, \Ref{cs}.

Likewise, we shall bring the arising vector equations into the form
\Ref{3dvec}. This is accomplished by noting that the equation
\bea
D^\mu \left( z^{\gamma}\,F_{\mu\nu}\right) &=&
\ft{1}{\tilde{\ell}^2}\,\lambda\,z^{\gamma}\,A_\nu \;.
\la{3dvecM}
\eea
on pure $AdS_3$ may be transformed into \Ref{3dvec} by means of the
rescaling
\bea
A_\mu &\rightarrow& z^{\frac12(1+\gamma)}\left(A_\mu +
\ft{1+\gamma}{p^2 z}\,(\dd_\mu A_z - A_z'\delta^z_\mu ) \right) \;,
\qquad \lambda \rightarrow \lambda\mis1\pls\gamma^2\;.
\la{travec}
\eea
In addition, we shall encounter  another type of vector field
equations in the IIB reduction, namely the first order equations
\bea
F_{\mu\nu} &=& m \,\omega_{\mu\nu\rho}\,A^\rho \;,
\la{3dvec1}
\eea
on pure $AdS_3$. Iterating this equation leads to \Ref{3dvecM} with
$\gamma=0$, i.e.\ upon rescaling according to \Ref{travec} it induces
the standard form \Ref{3dvec} with
\be
E_0=\ft32+\ft12|m| \;.
\label{self}
\ee
Note that unlike \Ref{3dvec}, equation \Ref{3dvec1} gives rise to only
one physical degree of freedom.

\mathon
\subsection{Scalar sphere harmonics}
\mathoff

Let us first consider fluctuations containing the scalar sphere
harmonics $Y^{(k)}$, associated with the representation $(k000)$ and
satisfying
\bea
\square_{S^7} Y^{(k)} &=& -\ft14\,k(k+6)\,Y^{(k)}\;.
\eea
The extra factor of ${1\over 4}$ on the r.h.s./ comes from the $S^7$
radius $\tilde{\ell}^2=4\, \ell^2$ while $\ell^2$ will be set to one
throughout this section. The scalar sphere harmonics $Y^{(k)}$ appear
in the expansion of the following fluctuations
\bea
g_{\mu\nu} &=& \mathring{g}_{\mu\nu} \left(1+
h_3\,Y^{(k)}\right)+ h_{\mu\nu}\,Y^{(k)} \;, \qquad
\mathring{g}{}^{\mu\nu}h_{\mu\nu} ~=~ 0\;,\nn\\
g_{mn} &=& \mathring{g}_{mn} \left(1+ h_7\,Y^{(k)}\right) \;,\nn\\
\tau &=&  i e^{-\mathring{\phi}} - i e^{-\mathring{\phi}} \,(
i\chi + \varphi )\,Y^{(k)}\;,\nn\\
A_{\mu\nu} &=& \mathring{A}{}_{\mu\nu} + \left( b_{\mu\nu}+
i\,c_{\mu\nu}\right) Y^{(k)} \;,
\eea
The Einstein field equations imply $3h_3 + 5h_7=0$ which we shall use
to eliminate $h_3$. The $[\mu m]$ component of the three-form
equations of motion imply the following reduction
\bea 
b_{\mu\nu} &=& z^{-1} \omega_{\mu\nu\rho} \,\dd^\rho  b\;, 
\qquad
c_{\mu\nu} ~=~ z^2 \omega_{\mu\nu\rho} \, \dd^\rho \left(
z^{-\frac32} c \right) \;.
\eea
The rescaling of $c$ turns out to be convenient in the following.  The
fluctuation equations for $\chi$ and $c$ decouple from the others and
after some computation give rise to
\bea z^2 \chi_1'' -2z \chi_1' + \ft14(9-k^2) \chi_1 + p^2 z^2
\chi_1 &=& 0 \;,\non 
z^2 \chi_2'' -2z \chi_2' - \ft14(27+12k+k^2)
\chi_2 + p^2 z^2 \chi_2 &=& 0 \;,
\eea
for the combinations
\bea \chi_1 &=& \chi + \ft16(6+k)\, c \;,\qquad \chi_2 ~=~ \chi
-\ft16 k\, c \;. \eea
These equations are indeed of the scalar type (\ref{34sca}) with
energies
\bea E^{(1)}_0 &=& \ft12k+\ft32 \;,\qquad
E^{(2)}_0 ~=~ \ft12k + \ft92 \;.
\eea
Comparing to table~\ref{specIIB} we hence find precise agreement with
the scalars in the second and fourth column --- note that $k=n\pm1$.
The remaining three scalar fields ($\varphi$, $b$, $h_7$) are somewhat
more difficult to analyze, as they mix with the fluctuations of the
three-dimensional metric. It turns out, that the combinations
\bea \phi_1 &=& 6\,h_7+\varphi  -\ft13\,(6+k)\, b\;,  \qquad
\phi_2 ~=~ 6\,h_7+\varphi  +\ft13\,k\,b \;,\label{p12} \eea
satisfy separate equations of motion
\bea z^2 \phi_1'' -2z \phi_1' + \ft14k(6-k)\, \phi_1 + p^2 z^2
\phi_1 &=& 0 \;, \non 
z^2 \phi_2'' -2z \phi_2' -
\ft14(72+18k+k^2)\,\phi_2 + p^2 z^2 \phi_2 &=& 0 \;,\eea
which again are equations of the type \Ref{34sca} with energies
\bea
E^{(1)}_0 &=& \ft12k \;,\qquad E^{(2)}_0 ~=~ \ft12k + 6 \;. \;.
\eea
This corresponds to the scalars in tables~5, and~9, i.e.\ the first
and last column of table~\ref{specIIB} ($k=n\pm2$).  The fifth scalar
in this sector is more complicated to identify. At momentum ${\bf
p}=0$, we find that the combination
\ben
\phi_3  ~=~
6(16+k(k\pls6))\, h_7 - 3(6+k(k\pls6))\,\varphi + k(k\pls6)\,b +
2z\left(18h_7'+3\varphi'-4b'\right) \;,
\een
satisfies scalar field equations \Ref{34sca} with energy $\ft12k\pls
3$ in agreement with the third column of table~\ref{specIIB}, cf.\
table~7.

\mathon
\subsection{Vector harmonics}
\mathoff

We consider now fluctuations carrying an $SO(7)$ vector index, i.e.\
the scalar sphere harmonics $Y^{(k)}_m$ associated with the
representation $(k\mis1100)$ and satisfying
\be
\square_{S^7}\,Y^{(k)}_m=
-\ft14\left[k(k+6)-1\right]\,Y^{(k)}_m\;,\qquad D^m\,
Y^{(k)}_m=0 \;.
\label{y1}
\ee
They appear in the field fluctuations
\bea
A_{\mu\nu\sigma m} &=& c_{\mu\nu\sigma}\, Y^{(k)}_m\;,\nn\\
A_{\mu m} &=& (b_{\mu}+i\,c_{\mu})\, Y^{(k)}_m\;,\nn\\
h_{\mu m} &=& h_{\mu}\, Y^{(k)}_m \;.
\label{ansatzV}
\eea
The self-duality equations for the five-form
\bea
F_{\mu\nu\sigma m_1 m_2}=2\, \partial_{[m_1}\,
Y_{m_2]}\, \left(c_{\mu\nu\sigma}+\ft38\, \mathring{A}_{[\mu\nu}\,
c_{\sigma]} \right)&=&0 \;,
\nn
\eea
can be used to determine $c_{\mu\nu\sigma}$ in terms of
$c_\sigma$. The remaining equations split as before into real and
imaginary parts. We focus on the fields in the second and fourth
column of table~\ref{specIIB}, i.e.\ those coming from the imaginary
part. By plugging the ansatz (\ref{ansatzV}) into the three-form
equation in (\ref{eqIIB}), one finds
\ben
\hat{D}_P\,
\left(z^{-\frac34}\, \hat{F}^{P\mu n} \right)-
z^{-\frac34}\,\mathring{P}{}_z\, \hat{F}^{z\mu
n}=0
\een
with
\bea
F_{m \mu n}&=&2\,\partial_{[n}Y_{m]}\, c_\mu\;,
\qquad
F_{\nu \mu n}~=~2\,\partial_{[\nu}\,c_{\mu]}\,Y^{n}\;.
\nn
\eea
Using \Ref{Dback}, this equation may be rewritten in pure $AdS_3$
background as
\bea
D^\nu\, \left( z^{-2}\, F_{\nu\mu}
\right)\,Y^n+z^{-2}\, c_\mu\,\left(\square_{S^7}-\ft32\right) Y^n
&=& 0 \;,
\eea
which reduces to a vector equation of type (\ref{3dvecM}) with
$\gamma=-2$, $\lambda=\ft14(k^2+6\,k+5)$. Upon rescaling according to
\Ref{travec}, this gives rise to the standard vector equation
\Ref{3dvec} with energy
\be
E_0=\ft12 k+ 3 \;.
\ee
This reproduces the values in tables~6 and~8 ($k=n$) in this sector. A
similar analysis of the real part of the fluctuation equations
corresponds to the vector fields in the odd columns of
table~\ref{specIIB} and yields the energies
\be
E^{(1)}_0=\ft12 k + \ft32 \;,\qquad E^{(2)}_0 =\ft12 k +\ft92\;,
\la{Evec15}
\ee
with $k=n\pm1$. In particular, for the $SO(8)$ gauge vector fields
arising at $n=0$, this gives $\delta=\ft12$, cf.~\Ref{cha}. Computing
the associated current-current correlation functions according
to~\Ref{cd}, we hence recover the result of~\cite{Gherghetta:2001iv}.

\mathon
\subsection{Two-form harmonics}
\mathoff

Next, we analyze the fluctuations involving the rank two form
$Y^{(k)}_{mn}$ on $S^7$ associated to the representation
$(k\mis1011)$:
\be
3\,D^m\, D^{\vphantom{l}}_{[m}\,Y_{np]}^{(k)}
=-\ft14(k^2+6k+8)\, Y^{(k)}_{np}
\equiv \lambda \, Y^{(k)}_{np} \;,
\qquad D^mY_{mn} = 0\;.
\label{y2}
\ee
They appear as
\bea
A_{m n} &=& (b+i\, c_1)\,Y_{m n} \;, \nn\\
A_{\mu\nu m n} &=& c_{\mu\nu} \, Y_{m n} \;, \nn\\
A_{m_1\dots m_4}&=&{c_2\over 3!}\,
\hat{\omega}_{m_1\dots m_4}{}^{m_5\dots m_7}\,
\partial_{m_5}\,Y_{m_6 m_7} \;.
\label{ansYmn}
\eea
Plugging (\ref{ansYmn}) into the five-form self-duality equations
we find
\bea
\ft13\,\lambda \,c_2\, \omega_{\mu\nu\rho}&=&
3\,\partial_{[\rho}\,
C_{\mu\nu]}-\ft14 \mathring{H}_{\mu\nu\rho}\,c_1 \;, \nn\\[1ex]
C_{\mu\nu} &=&-\ft13\,\omega_{\mu\nu\rho}\partial^{\rho}\,c_2 \;,
\label{5mn}
\eea
with
\ben
C_{\mu\nu}\equiv c_{\mu\nu}+\ft18 \mathring{A}_{\mu\nu}\,c_1 \;,
\een
and $\lambda$ defined through (\ref{y2}). The imaginary part of the
three-form equation
\ben
\hat{D}^P\,
G_{mnp}+\mathring{P}{}^z \, G_{mnz}^*=-\ft29\,i\,\lambda \,c_2\,
\hat{\omega}^{\mu\nu\rho}\, \mathring{H}_{\mu\nu\rho}\, Y_{mn} \;,
\een
reduces to
\be z^5\,\partial_\mu\left( z^{-3}\partial^{\mu}\,
c_1\right)+ \lambda\, c_1-4\,\lambda z\,c_2\,=0 \;.
\label{2mn}
\ee
Equation (\ref{5mn}) can be used to solve for $C_{\mu\nu}$ in terms of
$c_2$. Combining with (\ref{2mn}), we are left with the following
system of second order differential equations:
\bea
z^2\,\bar{c_2}'' -2\, z\,\bar{c_2}'+p^2\, z^2\, \bar{c_2}
-\ft14( k(k\pls6)+3)\, \bar{c_2}+\ft94\,\bar{c_1}&=&0 \;,
\nn\\
z^2\,\bar{c_1}'' -2\, z\,\bar{c_1}'+p^2\, z^2\, \bar{c_1} -\ft14(
k(k\pls6)+15)\, \bar{c_1} +(k(k\pls6)+8) \,\bar{c_2}&=&0 \;,
\eea
with $\bar{c_1}=z^{-{1\over 2}}\, c_1$, $\bar{c_2}=z^{1\over 2}\,
c_2$. This rescaling is required to bring the equations into the
standard form \Ref{34sca}. After straightforward diagonalization they
give rise to differential equations of the standard type \Ref{diffeq},
(\ref{cs}) with energies
\bea E^-_0 &=& \ft12k+\ft32 \;,\qquad
E^+_0 ~=~ \ft12k +\ft92 \;.\label{emn} \eea
These energies match the values in tables~6,~8 ($k=n\pm1$).

\mathon
\subsection{Three-form harmonics}
\mathoff

Finally, we consider fluctuations of the self-dual five form carrying
the rank three sphere harmonics $Y^{(k\pm)}_{mnp}$ satisfying
\be
\ft{1}{3!}\, \omega_{m_1\dots m_7}\, D^{m_1}\,
Y^{(k\pm)\,m_2 m_3 m_4}=\pm \ft12(k+3)\, Y^{(k\pm)}_{m_5 m_6 m_7} \;,
\ee
with the sign $\pm$ distinguishing the two representations
$(k\mis1020)$ and $(k\mis1002)$.

Inspection of table~\ref{specIIB} suggests that these fluctuations
give rise to the vector fields appearing in the second and fourth
column. Recall that the vector fields~\Ref{ansatzV} satisfy the
standard equation \Ref{3dvec} giving rise to two physical degrees of
freedom. In contrast, the representations $(k\mis1020)$, $(k\mis1002)$
each appear only once in this table. This already suggests that these
fields rather satisfy a first order equation which indeed will turn
out to be the case. Consider the ansatz
\bea
A_{\mu m n p} &=& c_{\mu}\, Y^{(k\pm)}_{m n p} \;, \qquad
A_{m n p q} ~=~ c \, D_{[m}Y^{(k\pm)}_{n p q]} \;.
\eea
The $[m_1\dots m_5]$ component of the five-form self-duality equations
implies $c=0$, while the $[\mu\nu m_1 m_2 m_3]$ component reduces to
the form~\Ref{3dvec1}
\bea
2\,
\partial_{[\mu}\, c_{\nu]}\,Y^{(k\pm)}_{m n p}&=&
\pm \ft12(k+3)\, \omega_{\mu\nu\rho}\,
c^\rho\,Y^{(k\pm)}_{m n p} \;.
\eea
According to our discussion above, this gives rise to fields with
energy
\be
E_0^{\pm}=\ft12 k+3 \;,
\ee
in agreement with the values in tables~6,~8 ($k=n$).

\mathon
\section{Type I supergravity on warped $AdS_3\times S^7$}
\mathoff

In this section, we consider the reduction of type I supergravity on
warped $AdS_3\times S^7$. The starting point is now given by the
decomposition of the ten-dimensional $({\bf 8_v}+{\bf 8_s})({\bf
8_v}+n_v\cdot {\bf 1})$ (with $n_v$ the number of vector multiplets)
field content under $SO(7)$:
\be
R_{SO(7)}^{\rm I}= (2\pls n_v) \cdot {\bf 1}
+(2\pls n_v)\cdot {\bf 7}+(2\pls n_v)\cdot {\bf 8}
+{\bf 21}+{\bf 27}+ {\bf 48} \;.
\label{so7I}
\ee
Combining this with table \ref{SO78} one finds:
\bea {\cal H}_I=
 \sum_{n=0}^{\infty} ({\bf 8_v}+n_v\cdot {\bf 1}){\bf (n000)}_{s}
\;,
\label{sumI}
\eea
with the multiplet ${\bf (n000)}_{s}$ given in \Ref{multI} above. The
entire spectrum is collected in table~10. The quantum numbers
for states sitting in the one of the first three columns can be read
off directly from those in type IIB computed above. In the rest of
this section we shall hence analyze the remaining last column which
collects the states coming from the gauge sector of type I.

\begin{table}[bt]
\centering
\begin{tabular}{||l|l|l|l||}
\hline \hline \footnotesize $(n\plss2000)$   & \footnotesize
$(n100)$   &
\footnotesize $(n000)$ &
\footnotesize $n_v \cdot(n\plss1000)$ \\
\footnotesize $(n\pls1010)$  & \footnotesize $(n001)\pls(n\miss1
110)$   &
\footnotesize $(n\miss1 010)$& \footnotesize  $n_v \cdot(n 010)$\\
\footnotesize $(n100)$  & \footnotesize $(n000)\pls(n\miss1
011)\pls(n\miss2 200)$   &
\footnotesize $(n\miss2 100)$ & \footnotesize$n_v \cdot(n\miss1 100)$\\
\footnotesize $(n001)$  & \footnotesize $(n\miss1 010)\pls(n\miss2
101)$   &
\footnotesize $(n\miss2 001)$ & \footnotesize$n_v \cdot(n\miss1 001)$\\
\footnotesize $(n000)$   & \footnotesize $(n\miss2 100)$    &
\footnotesize $(n\miss2 000)$ & \footnotesize$n_v \cdot(n\miss1 000)$\\
\hline \hline
\end{tabular}
\caption{\small Spectrum of I supergravity on $S^7$. }
\label{specI}
\end{table}

The linearized bosonic equations in the type I gauge sector read
\be \hat{D}_M\,( e^{\mathring{\phi}\over 2}\, \hat{F}^{MN})-\ft12
\, e^{\mathring{\phi}}\, \mathring{H}_{MNP}\, \hat{F}^{NP}=0 \;,
\label{typeIeq}
\ee
with $F_{MN}=2\,\partial_{[M}\, A_{N]}$\,.
Plugging in the vacuum solution (\ref{vacuum}) one finds
\bea
D_N\, F^{Nm}&=&
\left(\square_{AdS}+\square_{S^7}-\ft32\right)\, A^m~=~0 \;,
\nn\\
D_N\,F^{N\mu}-\ft32\,\omega^{\mu\nu\rho}\, F_{\nu\rho}
&=&D_\nu\,F^{\nu\mu}+\square_{S^7}\,
A^{\mu}-\ft32\,\omega^{\mu\nu\rho}\, F_{\nu\rho}~=~0 \label{eq1}
\;.
\eea
Remarkably, all the explicit $z$-dependence in these equations has
been canceled against the non-trivial dependence of the dilaton, three
form and warped metric. The gauge theory effectively lives on pure
$AdS_3\times S^7$. Consistently, the type I gauge multiplet (the last
column of table (\ref{specI}), cf.~table~10) matches the
structure of short multiplets of $OSp(8|,2,\bbR)_R\times SL(2,\bbR)_L$
found in \cite{Gunaydin:1986fe}. We will hence in this sector bring
all equations into the pure $AdS$ form. Scalar fields for example will
be normalized such as to satisfy~\Ref{diffeq}, \Ref{cs} with $a=1$
rather than $a=\ft32$, etc.

The first equation in \Ref{eq1} for the three-dimensional scalars
$A_m$ can be solved in terms of the ansatz
\bea
A_m&=&a\, Y_m^{(k)} \;,
\eea
with $Y^{(k)}_m$ associated to the representation $(k\mis1100)$ and
satisfying (\ref{y1}). Plugging in (\ref{eq1}) we are left with an
equation of scalar type (\ref{diffeq}) with $a=1$ and
\bea
E_0&=& \ft12 k+\ft52 \;,
\eea
which matches the result for the scalar component in table~10.
The equations of motion for the three-dimensional vector components
on the other hand can be solved by means of the ansatz
\be A_\mu=e_\mu^a\, a_{a}\,Y^{(k)} \;.
\ee
They reduce to two sets of first order vector equations \Ref{3dvec1}
with
\be
m_{\pm}=\ft32\pm \ft12(k+3) \;.
\ee
Energies can be read off from (\ref{self}) and are given by
\bea E_0^- &=&\ft12 k +1 \;,\qquad
E^+_0 ~=~\ft12 k+4 \;.
\label{apm}
\eea
Recalling that $Y^{(k)}$ is associated to the representation $(k000)$
we see that these energies match the values quoted in
table~10.

\section{Chiral primaries in the gauge theories}

In the previous sections we have determined the spectrum of masses and
charges of single particle Kaluza-Klein states in the reduction of
ten-dimensional supergravities on warped $AdS_3\times S^7$. The aim of
this section is to compare these results to their dual descriptions in
terms of primary operators in the boundary theories.

The nature of the dual theory is substantially different depending on
whether we consider systems of fundamental or D-strings. In the former
case, states in a floor of the Kaluza Klein tower below level $N$ are
associated to fundamental string states with charge $N$. The physics
at finite $g_s$ is presumably described by deformations of the
two-dimensional ${\cal N}=(8,8)$ and ${\cal N}=(8,0)$ sigma models
associated to ten-dimensional strings on flat spacetimes. Systems
involving fundamental strings are particularly interesting since due
to the absence of RR backgrounds string theory can be handled in a
better controlled way.  In the D-string case the dual gauge theory is
defined via quantization of the lowest open string modes governing the
low energy dynamics of $N$ nearby D-strings. They result into
effective $U(N)$ and $SO(N)$ two-dimensional gauge theories for the
type II and type I D-string, respectively. We focus our study on these
D-string systems.

Let us start by considering the D-string system in type IIB. The
effective $U(N)$ gauge theory describing the low energy dynamics of
$N$ nearby D-strings is obtained from dimensional reduction of ${\cal
N}=1$ SYM in $D=10$ down to two dimensions. The field content
comprises (besides the gauge vector field) eight adjoint scalars
$\phi^I$, and left and right moving fermions $S^a$, $S^{\dot{a}}$,
transforming in the ${\bf 8_v}$, ${\bf 8_s}$, and ${\bf 8_c}$,
respectively, of the $SO(8)$ R-symmetry group.  The theory is
manifestly invariant under sixteen supersymmetries. The Poincar\'e
symmetry group $ISO(1,1)\times SO(8)$ and supersymmetries match the
isometries and Killing spinors of the $AdS_3\times S^7$ background
(\ref{vacuum}), as expected.

The analysis of primary operators follows straightforwardly from that
of ${\cal N}=4$ SYM from which the two-dimensional gauge theory
descends upon dimensional reduction. As in the four-dimensional case,
chiral primaries are associated to completely symmetrized, traceless
operators $\CO_m\equiv{\rm Tr} \left(\phi^{I_1}\dots
\phi^{I_m}\right)$, $m=2, 3, \dots$, built from scalar fields in the
${\bf 8_v}$ and transforming in the $(m000)$ representation of the
$SO(8)$ R-symmetry group.  The missing of the $m=1$ state is due the
fact that $\phi^I$ is an $SU(N)$ rather than a $U(N)$ matrix. The
remaining primaries can be found by acting with the fermionic charges
$Q^a, \tilde{Q}^{\dot{a}}$ on the chiral primary. Half of these
charges $Q^{a}$ realize the eight on-shell supersymmetries and span
the columns of table \ref{specIIB}.  The remaining
$\tilde{Q}^{\dot{a}}$ act horizontally between the various
columns. The state at the bottom right is reached by $Q^4\,\tilde{Q}^4
\CO_m= {\rm Tr}\, \left({\cal F}^4\,\phi^{I_1}\dots
\phi^{I_{m-4}}\right)$.  The resulting spectrum of $SO(8)$
representations and spins is listed in tables~5--9 in the Appendix
($m=n\pls2$). The results agree with those coming from the KK harmonic
analysis in the last section.  The $E_0$ quantum number can also be
related to conformal dimensions in the parent SYM in $D=4$. Recalling
that $E_0=\ft12\Delta$, we see that the chiral primary $\CO_m$ carries
a total energy $E_0=\ft12 m$ in agreement with the supergravity
result. Notice that this result generalizes straightforwardly to
arbitrary dimensions: chiral primaries at the floor $m$ in the KK
reductions on $AdS_{d+1}\times S^{9-d}$ will carry energies
$E_0=\ft{\ell}{\tilde{\ell}}\,m={(6-d)\over 2}\,m$ in agreement with
(\ref{u0}).

It is interesting to compare the above results with the ${\cal
N}=(8,8)$ SCFT describing the infrared physics.  Although the
supergravity picture break down in this regime, one can still hope
that protected quantities like the spectrum of chiral primaries, and
their two- or three-point functions are preserved under the flow
towards the IR. According to \cite{Dijkgraaf:1996xw}, chiral primaries
in a symmetric product SCFT $M^N/S_N$ are counted by an elliptic index
which can be thought as a second quantized partition function for the
non-trivial cohomologies of $M$. In our case, this leads to a single
generator associated to $h_{0,0}=1$ in $\bbR^8$. This is in agreement
with the results in the supergravity side where a single ``${\cal
N}=(8,8)$ chiral primary''\footnote{By this, we mean a state at the
top left in table~\ref{specIIB}. In the conformal limit all states in
this table should come together to build a multiplet of the maximal
${\cal N}=(8,8)$ supersymmetry.}  was found at each level $m$ in the
KK tower.  Despite this encouraging evidence of a correspondence
between the spectra of SCFT/supergravity chiral primaries, the
mismatch between the supergroups in the SCFT and supergravity regimes
still obscures the precise dictionary between masses and charges.  It
would be nice to perform a systematic study of the SCFT/supergravity
elliptic genera along the lines of \cite{deboer,ghmn} to put this
correspondence on more firm grounds.

Finally, let us comment on the $n=-1$ states in table
\ref{specIIB}. These states, missing in the gauge/supergravity
spectrum are the so called singletons and typically associated
with global degrees of freedom living at the $AdS$ boundary. In the
present situation, they include ${\bf 8}_v$ scalars and ${\bf
8}_s,{\bf 8}_c$, spin-$\ft12$ fields which match the worldsheet
content of the dual IR CFT. In addition, there is an extra scalar
at $n=-1$ in table~6 which transforms as a singlet under $SO(8)$.
This state corresponds to a rather special operator in the dual
gauge theory. From table~\ref{specIIB} we see that this mode is
associated to the first irrelevant operator invariant under the
whole ${\cal N}=(8,8)$ supersymmetry. These are precisely the
requisites met by the DVV $\bbZ_2$-twisted operator
\cite{Dijkgraaf:1997vv} associated to deformations of the IR SCFT
away from the $(\bbR^8)^N/S_N$ orbifold point. Its dimension $\Delta=2
E_0=3$ and spin $s_0=0$ also match those of the DVV operator in
\cite{Dijkgraaf:1997vv}.

The above analysis can be easily extended to the type I case.  The
D-string bound state dynamics is now described by a $O(N)$ gauge
theory with eight scalars $\phi^I$ and right moving fermions $S^a$ in
the symmetric representation of $O(N)$, eight left moving adjoint
fermions $S^{\dot{a}}$, and 32 left moving $\lambda^i$ fermions
transforming as singlets of $SO(8)$ and bifundamentals of $O(N)\times
SO(32)$. The different $O(N)$ representations of left and right moving
fermions compensate for the relative sign in the action of the
$\Omega$-projection on type IIB D-string fields. The surviving
supersymmetry is ${\cal N}=(8,0)$.  The extra fundamentals arise from
quantization of D1-D9 open strings. Chiral primaries associated to
bulk supergravity modes can be derived from those in type IIB theory
via $\Omega$-projection. In order to do this we first decompose the
${\cal N}=(8,8)$ supermultiplets of table~\ref{specIIB} in terms of
${\cal N}=(8,0)$ as given by tables 5--9, collected in the Appendix;
each of them has an ${\cal N}=(8,0)$ chiral primary on the
top. Operators sitting in tables~5,~7,~9, are kept by the
$\Omega$-projection and the $SO(8)$, spin and scaling quantum numbers
follow from their type IIB cousins. From the gauge theory point of
view, the projection of operators in the second and fourth columns of
table~\ref{specIIB} can be easily understood since they are associated
to traces involving an odd number of antisymmetric matrices. In
addition, we have $n_v={\rm dim}\left[SO(32)\right]$ extra ${\cal
N}=(8,0)$ multiplets descending from the gauge chiral primary ${\rm
Tr}\, \left(\lambda^{[i} \, \lambda^{j]}\, \phi^{I_1}\dots
\phi^{I_{m-1}}\right)$ in the $(m-1000)$ representation of
$SO(8)$. This agree with the supergravity spectrum in table~10.
The energies $E_0$ also match the values in the table with an
extra contribution of $1$ in the gauge sector coming from
$\lambda^{[i}\, \lambda^{j]}$.

In the infrared limit the gauge theory flows to a two-dimensional
SCFT given in terms of $N$ copies of heterotic strings moving on
$(\bbR^8)^N/S_N$ \cite{Gava:1998sv}.  The singletons at $n=-1$ now
include, besides the ${\bf 8_v}$ scalars and ${\bf 8_s}$ right
moving fermions, the $SO(8)$ singlets $\lambda_L^{[i}
\,\lambda_L^{j]}$ in the adjoint of the $SO(32)$ D9-gauge group in
agreement, with the SCFT worldsheet content. The spectrum of ${\cal
N}=(8,0)$ chiral primaries in the IR SCFT is now associated to a
second quantized partition function with $n_v+8$ bosonic generators,
coming from chiral primaries at the massless level of a single copy of
the heterotic string on $\bbR^8$.  Upon compactifications on a circle
the $SO(32)$ heterotic gauge group is generically broken to
$U(1)^{16}$ giving $n_v=16$.  This agrees with the supergravity
results above, with $n_v+8$ the number of (single-particle) chiral
primaries found at each level in the KK tower.

\section{Conclusions}

There are several interesting questions raised by our study. 
As has been remarked in \cite{Boonstra:1998mp}, the vacuum
configuration \Ref{vacuum} implies the existence of gauged
supergravities which do not admit a ground state but rather a domain
wall solution. The field content of these theories is given by the
supermultiplets ${\bf(0000)}_{\rm IIA}$ and ${\bf(0000)}_{\rm IIB}$,
in tables~\ref{specIIA} and \ref{specIIB}, respectively. Whereas the
former theory may just be obtained from the circle reduction of the
$SO(8)$ gauged four-dimensional theory of~\cite{deWit:1982ig}, the
latter theory is as yet unknown. We suspect that both these theories
may in fact be embedded into the maximal gauged three-dimensional
supergravity constructed in~\cite{ns} which admits an ${\cal N}=(8,8)$
$AdS_3$ vacuum. It would be interesting to understand within this
theory the flow of the matrix string models to the orbifold SCFT's
with target space $(\bbR^8)^N/S_N$ and enhanced supersymmetries in the
infrared limit. Domain wall solutions interpolating between
two-dimensional SCFT's have been previously studied in
\cite{Morales:2001zp}.

In \cite{Bonelli:1998wx}, the authors have shown that the string genus
expansion is reproduced by a simple counting of moduli in the matrix
string SCFT. It would be nice to understand the meaning of this result
from the supergravity perspective.

The analysis of domain wall/QFT duals in other dimensions is of
obvious interest. In~\cite{Hammou:1999in}, an exact CFT description of
the infrared dynamics of D5-branes in type I theory was proposed in
terms of second quantized type II strings on $(\bbR^4\times
K3)^N/S_N$. The study of this SCFT from the dual supergravity
perspective could provide further insights in the physics content of
these domain wall/QFT correspondences.

Another interesting direction is to use the $AdS_3\times S^7$ dual
description to explore the non-perturbative physics described by type
II and heterotic matrix string models.  Similarly, one could exploit
the $AdS_2\times S^8$ dual description of D0 matrix theories
\cite{Banks:1996vh} in order to study M-theory physics.  The
celebrated matrix theory description of 11d graviton-graviton
scattering in terms of ${\rm Tr}\,{\cal F}^4$ correlation function in
SYM in this perspective translates into a tadpole computation of the
corresponding dual supergravity field in the string background.  For
$d=2$, the associated supergravity field corresponds to the state at
the bottom-right of table \ref{specIIB} for $n=2$. We have identified
this field in (\ref{p12}) as a linear combination $\phi_2$ of the
s-waves ($k=0$) of the dilaton and the metric trace on $S^7$.
Remarkably, this particular linear combination does not involve the RR
$b$-field. It is interesting to compare this with the results of
\cite{Bianchi:2000vb}, where the ${\rm Tr}\, {\cal F}^4$ term in SYM
with 16 supercharges was shown to be reproduced by a tadpole
computation again purely in terms of NSNS fields.\footnote{We thank
M.~Bianchi for drawing our attention to this point} It would be nice
to have more quantitative tests of these ideas.

Finally, it is natural to investigate the pp-wave limits
\cite{Berenstein:2002jq} of the correspondences under study
here. Penrose limits of near horizon Dp-brane geometries have been
recently studied in \cite{Gimon:2002sf}.
We hope to come back to some of these issues in the near future.

\vspace*{.5cm}

\subsection*{Acknowledgements}

\noindent
We wish to thank J.~de Boer, B.~de Wit, M.~Trigiante and S.~Vandoren
for useful discussions. This work is partly supported by EU contract
HPRN-CT-2000-00122.

\mathon
\section*{Appendix: ${\cal N}=(8,0)$ short multiplets}
\mathoff

In this appendix we display the $ISO(1,1)\times SO(8)$ quantum
numbers and the structure of ${\cal N}=(8,0)$ supermultiplets
entering in KK reductions of ten-dimensional supergravities on
$S^7$. The ${\cal N}=(8,0)$ supersymmetry is chosen here as the
basic unit common to all five ten-dimensional strings in order to
present a unifying treatment. States in the reductions of type II
strings are naturally organized in terms of ``${\cal N}=(8,8)$
supermultiplets'' displayed in tables~\ref{specIIA},~\ref{specIIB}
which decompose into sixteen ${\cal N}=(8,0)$ supermultiplets
transforming as ${\bf 8_v}+{\bf 8_c}$ or ${\bf 8_v}+{\bf 8_s}$ of
the $SO(8)$ Lorentz group for type IIA and IIB, respectively. We
denote by $Q$ the supersymmetry charges acting inside of ${\cal
N}=(8,0)$ supermultiplets. These corresponds to the eight on-shell
supersymmetries.  Fermionic charges bringing us from one ${\cal
N}=(8,0)$ supermultiplet to another one inside a big "${\cal
N}=(8,8)$ multiplet" will be denoted by $\bar{Q}$($\tilde{Q}$) for
type IIA (IIB).

From a two-dimensional point of view, supersymmetry charges (as
two-dimensional fermions) transform naturally under $SO(8)_L\times
SO(8)_R$ but only the diagonal $SO(8)$ Lorentz group is preserved by
gauge interactions. The displayed $SO(8)$ quantum numbers will always
refer to this diagonal subgroup.  States will be labeled by the
$SO(8)$ Dynkin labels $(n_1,n_2,n_3,n_4)$ and $(\ell_0,\bar{\ell}_0)$
quantum numbers.  The eight supersymmetry charges $Q$ decompose under
$SO(2)\times SO(6)\sim U(1)_F\times SU(4)$ as four creation and four
annihilation with $U(1)_F$ charges $q=-{1\over 2}$ and $q={1\over 2}$,
respectively. The same holds for $\bar{Q},\tilde{Q}$.  The $SO(8)$
content of a given state can be entirely determined from its
$U(1)_F\times SU(4)$ quantum numbers via the
dictionary~\cite{Gunaydin:tc}
\bea
\left[(a_1,a_2,a_3);q \right]&\rightarrow&
\left(q+\ft12\,(n_3-n_1-n_2),n_2-n_3,n_1-n_2,n_3 \right)
\;,
\eea
with $(a_1,a_2,a_3)$ the numbers of boxes in each row of the Young
tableaux diagram characterizing the $SU(4)$ representation. In
addition, fermionic charges $Q$ raise by ${1\over 2}$ the value of
$\ell_0$, while $\bar{Q},\tilde{Q}$ raise that of $\bar{\ell}_0$.

Let us start by considering the type IIA spectrum. We start from
the highest weight state $| \, \Omega\, \rangle=|0\rangle$ with
$\left[(n+2000);\ft14(n+2),\ft14(n+2)\right]$ quantum numbers.
The states on the top of tables 5--9 are found by iterative
actions of $\bar{Q}$ on $|\Omega\rangle$. States in the same table
can be reached via $Q$ supersymmetry actions. The two set of
fermionic charges are in
the fundamental of $SU(4)$ and carry the following quantum numbers: %
\bea
Q:\quad\quad\left[(a_1,a_2,a_3),q;\ell_0,\bar{\ell}_0\right]
=\left[(1,0,0),-\ft12;0,\ft12\right]\nn\\
\bar{Q}:\quad\quad\left[(a_1,a_2,a_3),q;\ell_0,\bar{\ell}_0\right]
=\left[(1,0,0),-\ft12;\ft12,0\right]
\label{susy}
\eea
The remaining quantum numbers displayed in the tables
are the scaling dimensions $E_0$ and spins $s_0$ related to the
$(\ell_0,\bar{\ell}_0)$ through
\bea E_0&=&
\ell_0+\bar{\ell}_0\;,\qquad s_0~=~ \ell_0-\bar{\ell}_0\;.
\eea

The type IIB spectrum follows in the same way with the only
difference that the $\tilde{Q}$ supercharges carry now opposite
chirality, i.e.\
\be
\tilde{Q}:
\quad\quad\left[(a_1,a_2,a_3),q;\ell_0,\bar{\ell}_0\right]
=\left[(1,1,1),-\ft12;0,\ft12\right]
\label{susyB}
\ee
This gives rise to the $SO(8)$ contents displayed in tables~6
and~8. Tables 5, 7, 9 coincide with those for type IIA since the
highest weight state are now in real $SO(8)$ representations.
Finally, the spectrum of type I on $S^7$ can be obtained by
keeping the multiplets in tables~5,~7,~9, and adding $n_v$ gauge
multiplets with content given by table~10.

It is instructive to understand these results from the representation
theory of the group of conformal isometries $SL(2,\bbR)_L\times
OSp(8|2,\bbR)_R$. Its unitary representations have been constructed
in~\cite{Gunaydin:1986fe} on the Fock space of super-oscillators
\be {\bf \xi}^A_R = \left( {\bf a}^\dagger \atop {\bf \alpha}^i
\right)_R\;, \quad {\bf \eta}^A_R = \left( {\bf b}^\dagger \atop
{\bf \beta}^i \right)_R\;,\qquad {\bf a}^\dagger_L,~{\bf
b}^\dagger_L\;,\qquad i = 1, \dots, 4 \;, \ee
with $p={m\over 2}$ pairs of bosonic oscillators combined into vectors
${\bf a}_R$, ${\bf b}_R$, ${\bf a}_L$, ${\bf b}_L$ and similarly $p$
pairs of fermionic oscillators ${\bf \alpha}^i$, ${\bf \beta}^i$,
together transforming in the fundamental representation of $U(1|4)$
(for odd $m=2p+1$ one includes an additional super-oscillator). The
noncompact $SL(2,\bbR)_L\times SL(2,\bbR)_R$ subgroup is realized in
this formalism in terms of bilinears in the bosonic generators while
the compact $SO(8)$ is given in terms of fermionic ones.  Short
$SL(2,\bbR)_L\times OSp(8|2,\bbR)_R$ multiplets are constructed by
taking the vacuum $|0\rangle_{m}=|0\rangle_L\times|0\rangle_R$ as the
ground state and acting with creation operators. This corresponds to
table~5 (cf.~\cite{Gunaydin:1986fe}, table~11). Tables~6--9 in
contrast sit in $SL(2,\bbR)_L\times OSp(8|2,\bbR)_R$ long multiplets
obtained from the ground states $(a_L^\dagger)^s|0\rangle_L\times({\bf
\xi}_R^A)^s|0\rangle_R$, $s=1, \dots, 4$, however, these tables
show only the primary states satisfying the lower bound
$E_0\pls\ft12q=q_{\rm hws}$. E.g.\ the content of table~9 for $n=0$
corresponds to the single primary satisfying the bound in the long
multiplet given in table~9 of \cite{Gunaydin:1986fe}. Here, it
corresponds to a pure gauge mode of the graviton.

\bigskip

\begin{center}

\begin{tabular}{|c|c|c|c|c|}
\hline
  lowest state & $q$& $SO(8)$ & $E_0$ & $s_0$  \\
 \hline
  $|\Omega\,\rangle$ & $n+2$ & $(n\pls2000)$ & ${n\over 2}+1$ & $0$  \\
 $Q|\Omega\,\rangle$ & $n+{3\over 2}$ & $(n\pls1010)$ &
${n\over 2}+{3\over 2}$ & ${1\over 2}$  \\
 $Q^2|\Omega\rangle $ & $n+1$ & $(n100)$ & ${n\over 2}+2$ & $1$  \\
$Q^3|\Omega\rangle$
& $n+{1\over 2}$ & $(n001)$ & ${n\over 2}+{5\over 2}$ & ${3\over 2}$ \\
$Q^4|\Omega\rangle$ & $n$ & $(n000)$ & ${n\over 2}+3$ & $2$  \\
\hline
\end{tabular}
\begin{minipage}{15cm}
\smallskip

{\small Table 5: Supermultiplet with h.w.s.
$|\,\Omega\,\rangle=|0\rangle_{n+2}$, cf.~\cite{Gunaydin:1986fe},
table~11.  The $SO(8)$ content at $n=0$ is $35_v+56_s+28+8_s+1$.}
\end{minipage}

\bigskip
\bigskip
\bigskip
\bigskip

\begin{tabular}{|c|c|c|c|c|c|}
\hline
  lowest state & $q$& $SO(8)$ IIA & $SO(8)$ IIB &  $E_0$ & $s_0$  \\
 \hline
  $|\Omega\,\rangle$ & $n+{3\over 2}$ & $(n\pls1010)$ & $(n\pls1001)$ &
${n\over 2}+{3\over 2}$ & $-{1\over 2}$  \\
 $Q|\Omega\,\rangle$ & $n+1$ & $(n020)+(n100)$ & $(n\pls1000)+(n011)$ &
${n\over 2}+2$ & $0$   \\
 $Q^2|\Omega\rangle$ & $n+{1\over 2}$ & $(n001)+ (n\mis1110)$&
$(n010)+ (n\mis1101)$ & ${n\over 2}+{5\over 2}$ & ${1\over 2}$ \\
$Q^3|\Omega\rangle$ & $n$ & $(n000)+ (n\mis1011)$ & $(n\mis1100)+
(n\mis1002)$ & ${n\over 2}+3$ & $1$
\\
$Q^4|\Omega\rangle$ & $n-{1\over 2}$ & $(n\mis1010)$ & $(n\mis1001)$ &
${n\over 2}+{7\over 2}$ & ${3\over 2}$  \\
\hline
\end{tabular}
\begin{minipage}{15cm}
\smallskip

{\small Table 6: Supermultiplet with h.w.s.
$|\,\Omega\,\rangle=\bar{Q}|0\rangle_{n+2}=
a_L^\dagger|0\rangle_L\times{\bf\xi}_R^A|0\rangle_R$
($|\,\Omega\,\rangle=\tilde{Q}|0\rangle_{n+2}$) for type IIA (IIB).
The $SO(8)$ content at $n=0$ is
$56_s+35_c+28+8_s+1$\, ($56_c+8_v+56_v+8_c$). }
\end{minipage}

\bigskip

\begin{tabular}{|c|c|c|c|c|}
\hline
  lowest state & $q$& $SO(8)$ & $E_0$ & $s_0$  \\
 \hline
  $|\Omega\,\rangle$ & $n+1$ & $(n100)$ & ${n\over 2}+2$ & $-1$  \\
 $Q|\Omega\,\rangle$ & $n+{1\over 2}$ & $(n001)+(n\mis1110)$ &
${n\over 2}+{5\over 2}$ & $-{1\over 2}$  \\
 $Q^2|\Omega\rangle$ & $n$ & $(n000)+(n\mis1011)+(n\mis2200)$ &
${n\over 2}+3$ & $0$  \\
$Q^3|\Omega\rangle$& $n-{1\over 2}$ & $(n\mis2101)+ (n\mis1010)$ &
${n\over 2}+{7\over 2}$ & ${1\over 2}$ \\
$Q^4|\Omega\rangle$ & $n-1$ & $(n\mis2100)$ & ${n\over 2}+4$ & $1$ \\
\hline
\end{tabular}
\begin{minipage}{15cm}
\smallskip

{\small Table 7: Supermultiplet with h.w.s.
$|\,\Omega\,\rangle=\bar{Q}^2|0\rangle_{n+2} =
(a_L^\dagger)^2|0\rangle_L\times {\bf \xi}_R^A{\bf
\xi}_R^B|0\rangle_R$.  The $SO(8)$ content at $n=0$ is $28+8_s+1$ and
it is associated to gauge degrees of freedom coming from the massless
vector, gravitino and graviton respectively.}
\end{minipage}

\bigskip

\begin{tabular}{|c|c|c|c|c|c|}
\hline
  lowest state & $q$& $SO(8)$ IIA& $SO(8)$ IIB  & $E_0$ & $s_0$  \\
 \hline
  $|\Omega\,\rangle$ & $n+{1\over 2}$ & $(n001)$ & $(n010)$ &
${n\over 2}+{5\over 2}$ & $-{3\over 2}$  \\
 $Q|\Omega\,\rangle$ & $n$ & $(n000)+ (n\mis1011)$ &
$(n\mis1020)+(n\mis1100)$ & ${n\over 2}+3$ & $-1$  \\
$Q^2|\Omega\rangle$ & $n-{1\over 2}$ & $(n\mis1010)+ (n\mis2101)$ &
$(n\mis1001)+(n\mis2110)$ & ${n\over 2}+{7\over 2}$ & $-{1\over 2}$ \\
$Q^3|\Omega\rangle$& $n-1$ & $(n\mis2100)+ (n\mis2002)$  &
$(n\mis1000)+ (n\mis2011)$& ${n\over 2}+4$ & $0$ \\
$Q^4|\Omega\rangle$ & $n-{3\over 2}$ &$ (n\mis2001)$& $(n\mis2010)$ &
${n\over 2}+{9\over 2}$ & ${1\over 2}$  \\
\hline
\end{tabular}
\begin{minipage}{15cm}
\smallskip

{\small Table 8: Supermultiplet with h.w.s.
$|\,\Omega\,\rangle=\bar{Q}^3|0\rangle_{n+2}
=(a_L^\dagger)^3|0\rangle_L\times {\bf
\xi}_R^A{\bf \xi}_R^B{\bf \xi}_R^C|0\rangle_R$
($|\,\Omega\,\rangle=\tilde{Q}^3|0\rangle_{n+2}$) for type IIA (IIB).
The content at $n=0$ is pure gauge in the $SO(8)$ representations
$8_s+1$ ($8_c$).}
\end{minipage}

\bigskip

\begin{tabular}{|c|c|c|c|c|}
\hline
  lowest state & $q$& $SO(8)$ & $E_0$ & $s_0$  \\
 \hline
  $|\Omega\,\rangle$ & $n$ & $(n000)$ & ${n\over 2}+3$ & $-2$  \\
 $Q|\Omega\,\rangle$ & $n-{1\over 2}$ & $(n\mis1010)$ &
${n\over 2}+{7\over 2}$ & $-{3\over 2}$  \\
 $Q^2|\Omega\rangle$ & $n-1$ & $(n\mis2100)$ & ${n\over 2}+4$ & $-1$ \\
$Q^3|\Omega\rangle$& $n-{3\over 2}$ & $(n\mis2001)$ &
${n\over 2}+{9\over 2}$ & $-{1\over 2}$ \\
$Q^4|\Omega\rangle$ & $n-2$ & $(n\mis2000)$ & ${n\over 2}+5$ & $0$  \\
\hline
\end{tabular}
\begin{minipage}{15cm}
\smallskip

{\small Table 9: Supermultiplet with h.w.s.
$|\,\Omega\,\rangle=\bar{Q}^4|0\rangle_{n+2}
=(a_L^\dagger)^4|0\rangle_L \times{\bf \xi}_R^A{\bf \xi}_R^B{\bf
\xi}_R^C{\bf \xi}_R^D|0\rangle_R$.  There is a pure gauge
$SO(8)$ singlet at $n=0$.}
\end{minipage}

\bigskip

\begin{tabular}{|c|c|c|c|c|}
\hline
  lowest state & $q$& $SO(8)$ & $E_0$ & $s_0$  \\
 \hline
  $|\Omega\,\rangle$ & $n+1$ & $(n\pls1000)$ & ${n\over 2}+{3\over 2}$ &
$1$  \\
 $Q|\Omega\,\rangle$ & $n+{1\over 2}$ & $(n010)$ & ${n\over 2}+2$ &
${1\over 2}$  \\
 $Q^2|\Omega\rangle$ & $n$ & $(n\mis1100)$ & ${n\over 2}+{5\over 2}$ &
$0$  \\
$Q^3|\Omega\rangle$& $n-{1\over 2}$ & $(n\mis1001)$ & ${n\over 2}+3$ &
${1\over 2}$ \\
$Q^4|\Omega\rangle$ & $n-1$ & $(n\mis1000)$ & ${n\over 2}+{7\over 2}$ &
$1$  \\
\hline
\end{tabular}
\begin{minipage}{15cm}
\smallskip

{\small Table 10: Type I gauge supermultiplet with h.w.s.
$|\,\Omega\,\rangle=(a^\dagger)^2\,|0\rangle_L\times |0\rangle_R$. The
$SO(8)$ content at $n=0$ is $n_v\,(8_v+8_s)$.}
\end{minipage}

\end{center}

\end{document}